\documentclass[aps,prl,twocolumn,superscriptaddress,10pt,article,showpacs,longbibliography]{revtex4-2}
\usepackage{blindtext}
\usepackage{lipsum}
\usepackage{graphics}
\usepackage{amsmath}
\usepackage{graphicx}
\usepackage{graphics}
\usepackage{amssymb}
\usepackage{verbatim}
\usepackage{physics}
\usepackage{float}
\usepackage[normalem]{ulem}
\usepackage[colorlinks, citecolor=blue]{hyperref}
\usepackage[dvipsnames]{xcolor}
\usepackage{bbm}
\usepackage{enumitem}

\begin{document}

\title{A Universal Protocol for Quantum-Enhanced Sensing via Information Scrambling}

\author{Bryce Kobrin}
\thanks{These authors contributed equally to this work.}
\affiliation{Department of Physics, University of California Berkeley, Berkeley, CA 94720, USA}

\author{Thomas Schuster}
\thanks{These authors contributed equally to this work.}
\affiliation{Walter Burke Institute for Theoretical Physics and Institute for Quantum Information and Matter, California Institute of Technology, Pasadena, CA 91125 USA}
\affiliation{Department of Physics, University of California Berkeley, Berkeley, CA 94720, USA}

\author{Maxwell Block}
\affiliation{Department of Physics, Harvard University, Cambridge, MA 02138, USA}

\author{Weijie Wu}
\affiliation{Department of Physics, Harvard University, Cambridge, MA 02138, USA}

\author{Bradley Mitchell}
\affiliation{IBM Quantum, IBM Almaden Research Center, San Jose, CA 95120, USA}

\author{Emily Davis}
\affiliation{Department of Physics, Harvard University, Cambridge, MA 02138, USA}
\affiliation{Department of Physics,
New York University, New York, NY, 10003, USA}

\author{Norman Y.~Yao}
\affiliation{Department of Physics, Harvard University, Cambridge, MA 02138, USA}

\begin{abstract}
We introduce a novel protocol, which enables Heisenberg-limited quantum-enhanced sensing using the dynamics of any interacting many-body Hamiltonian.
Our approach---dubbed \emph{butterfly metrology}---utilizes a single application of forward and reverse time evolution to produce a coherent superposition of a ``scrambled'' and ``unscrambled'' quantum state.
In this way, we create  metrologically-useful long-range entanglement from generic local quantum interactions.
The sensitivity of butterfly metrology is given by a sum of local out-of-time-order correlators (OTOCs)---the prototypical diagnostic of quantum information scrambling.
Our approach broadens the landscape of platforms capable of performing quantum-enhanced metrology; as an example, we provide detailed blueprints and numerical studies demonstrating a route to scalable quantum-enhanced sensing in ensembles of solid-state spin defects.
\end{abstract}

\maketitle

Quantum-enhanced metrology leverages entanglement in a many-body system to improve the fundamental precision of sensing~\cite{giovannetti2006quantum,pezze2018quantum,giovannetti2011advances,degen2017quantum}.
The realization of this enhancement with large numbers of interacting particles represents an ongoing objective, with a wide range of potential applications including  atomic time-keeping~\cite{robinson2024direct,shaw2024multi,komar2014quantum}, gravitational wave sensing~\cite{schnabel2010quantum,aasi2013enhanced,tse2019quantum}, biological imaging~\cite{kucsko2013nanometre,wu2016diamond,bakhshandeh2022quantum,aslam2023quantum}, and the search for new fundamental physics~\cite{bradley2003microwave,zheng2016accelerating,aybas2021search,lehnert2022quantum}.

From a theoretical perspective,
the requirements for a many-body state to exhibit ``metrologically-useful''  entanglement are well understood~\cite{degen2017quantum,toth2014quantum,lu2015robust,frowis2016detecting,liu2020quantum,agarwal2022quantifying}.
However, from an experimental perspective, two overarching challenges limit the range of experimental platforms that can  realize this advantage: state preparation and signal readout. 
For the latter, the key tension is that any \emph{direct} measurement on a metrologically useful state will also be highly susceptible to noise in the read-out process~\cite{bohnet2014reduced,zhang2012collective}. 
To address this  challenge, seminal recent results have introduced sensing protocols that rely on \emph{time-reversed} dynamics~\cite{goldstein2011environment,davis2016approaching,frowis2016detecting,macri2016loschmidt}.
Crucially, these protocols exhibit significantly improved robustness to read-out noise and have been experimentally realized in the context of both spin squeezed~\cite{colombo2022time} and GHZ states~\cite{mooney2021generation}. 

\begin{figure}[t]
\includegraphics[width=\linewidth]{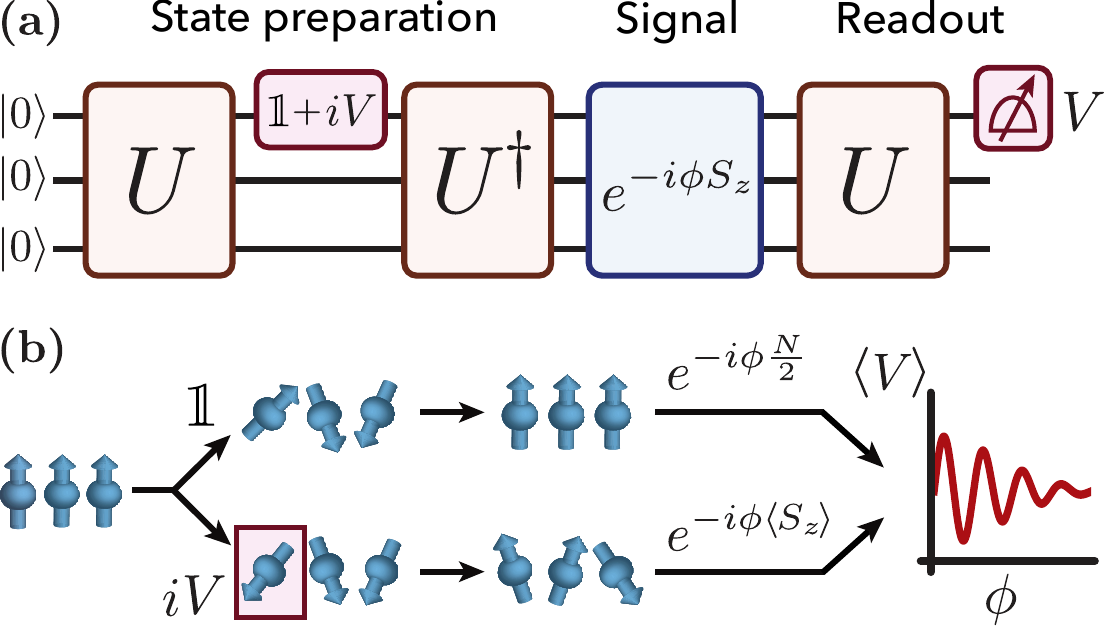}
\caption{\textbf{(a)} Schematic quantum circuit depicting our butterfly metrology protocol with local controls. The ``butterfly state'' is prepared via forward and reverse time evolution under a many-body Hamiltonian, $U = e^{-iHt}$, interleaved with a local rotation, $(\mathbbm{1} + i V)/\sqrt{2}$. The external signal, $e^{-i \phi S_z}$, is detected by evolving forward once more and measuring the local observable $V $. \textbf{(b)} In effect, the protocol performs interferometry between two quantum trajectories. In the first (top), the forward and reverse evolution cancel, yielding the polarized state $\ket{\textbf{0}}$. In the second (bottom), the perturbation $i V$ disrupts this cancellation, yielding the ``scrambled'' state, $i V(t) \ket{\textbf{0}}$. 
The two states acquire macroscopically different phases under the signal whenever $U$ is scrambling.
}
\label{fig1}
\end{figure}

For the former challenge, one of the most natural strategies to prepare a metrologically useful state is simply to perform Hamiltonian time evolution from an initial product state; however, the class of Hamiltonians for which this succeeds is extremely limited~\cite{degen2017quantum}.
Indeed, the only known examples consist of large-$S$ spin models~\cite{kitagawa1993squeezed,leroux2010implementation,lewis2018robust,zou2018beating,luo2017deterministic,lucke2011twin}, symmetry breaking evolution from a pure state~\cite{block2023universal}, and commuting central spin models~\cite{goldstein2011environment}.
This precisely encodes the challenge that while most quantum states are highly entangled,  only a vanishingly small subset can be utilized to perform enhanced sensing~\cite{hyllus2010not}.

In this Letter, we introduce a novel sensing protocol, dubbed ``butterfly metrology''.
Crucially, our protocol enables the preparation of metrologically-useful entangled states (from an initial product state) using completely \emph{generic} many-body Hamiltonians; in fact, our approach works for essentially any Hamiltonian, so long as it is not localized~\cite{abanin2019colloquium}.
The key insight underlying our approach is to use time reversal  not only to detect, but also to \emph{prepare} the metrological state (Fig.~\ref{fig1}). 
By doing so, our protocol generates a GHZ-like state from the dynamics of almost any  Hamiltonian.
This state is a coherent superposition of a fully polarized state and a \emph{scrambled} state [Fig.~\ref{fig1}(b)].
The latter has zero polarization on average, enabling our protocol to achieve sensitivities within a factor of two of the fundamental Heisenberg limit.

Our main results are threefold.
First, we establish a precise connection between butterfly metrology and quantum information scrambling \cite{fn_remark}.
In particular, we demonstrate that our protocol's sensitivity is exactly given by a sum of local out-of-time-order correlation functions (OTOCs)---the standard probe of scrambling dynamics~\cite{kitaev_simple_2015,shenker2014black,roberts2015localized}.
For fully scrambling dynamics, we immediately obtain a measurement sensitivity, $\eta$, that exhibits a Heisenberg scaling (i.e.~$\eta \approx 2/N$), with the number of particles. 
Second, we introduce a variant of our protocol that utilizes only \emph{global} control and readout, significantly reducing  the experimental requirements.
Finally, we highlight the broad applicability of our approach by providing detailed blueprints and numerical simulations for a variety of experimental platforms~\cite{supp}.
A particularly clear advantage over existing protocols is shown in the case of solid-state spin defects~\cite{awschalom2018quantum,taylor2008high,castelletto2023quantum}.
\emph{General strategy---}Consider an ensemble of $N$ spin-1/2 particles attempting to sense a weak external signal.
We imagine that the sensors are coupled to the signal, $\phi$, via the collective  spin operator $S_z = \frac 1 2 \sum_i \sigma^z_i$, where $ \sigma^z_i$ is a local Pauli operator acting on spin $i$.
A generic sensing protocol has three steps: (i) prepare a state, $\ket{\psi}$, on the $N$ spins, (ii) accumulate the signal, $e^{-i \phi S_z}$, and (iii) read-out an observable $M$ on the resulting state. 
The goal of any quantum sensing protocol is to optimize the sensitivity, $\eta = \Delta M_\phi/\partial_\phi \langle M \rangle_\phi$, of the measured expectation value, $\langle {M} \rangle_\phi$, to the signal $\phi$, where $\Delta M$ is the standard deviation of $M$.

The optimal sensitivity, over all possible observables, is lower-bounded by a property of the  state $\ket{\psi}$, namely, the so-called  quantum Fisher information (QFI)~\cite{braunstein1994statistical}.
For pure states, the QFI, $\mathcal{F}$, is simply given by twice the variance of the operator that couples to the signal, in our case, $S_z$.
This immediately yields a lower bound: $\eta \ge 1/\sqrt{\mathcal{F}} \equiv 1/(2 \Delta S_z)$. 

To provide a bit of intuition, let us consider how $\eta$ behaves in a few specific cases. 
For a product state, the standard deviation, $\Delta S_z$, is at most $\sqrt{N}$, implying a sensitivity scaling as, $\eta \sim 1/\sqrt{N}$, commonly known as the standard quantum limit (SQL). 
Most quantum states, despite being highly entangled, do not surpass this scaling. 
Indeed, a Haar-random state possesses, on average, the same QFI as an unentangled state~\cite{supp}.
On the other hand, the paradigmatic GHZ state, $(\ket{0}^{\otimes N} + \ket{1}^{\otimes N})/\sqrt{2}$, exhibits maximal correlations in the $z$-basis, and thus a standard deviation $\Delta S_z = N/2$ \cite{spin_basis}; this yields   a sensitivity, $\eta = 1/N$, known as the Heisenberg limit~\cite{bouwmeester1999observation}. 

\begin{figure}[t]
\includegraphics[width=\linewidth]{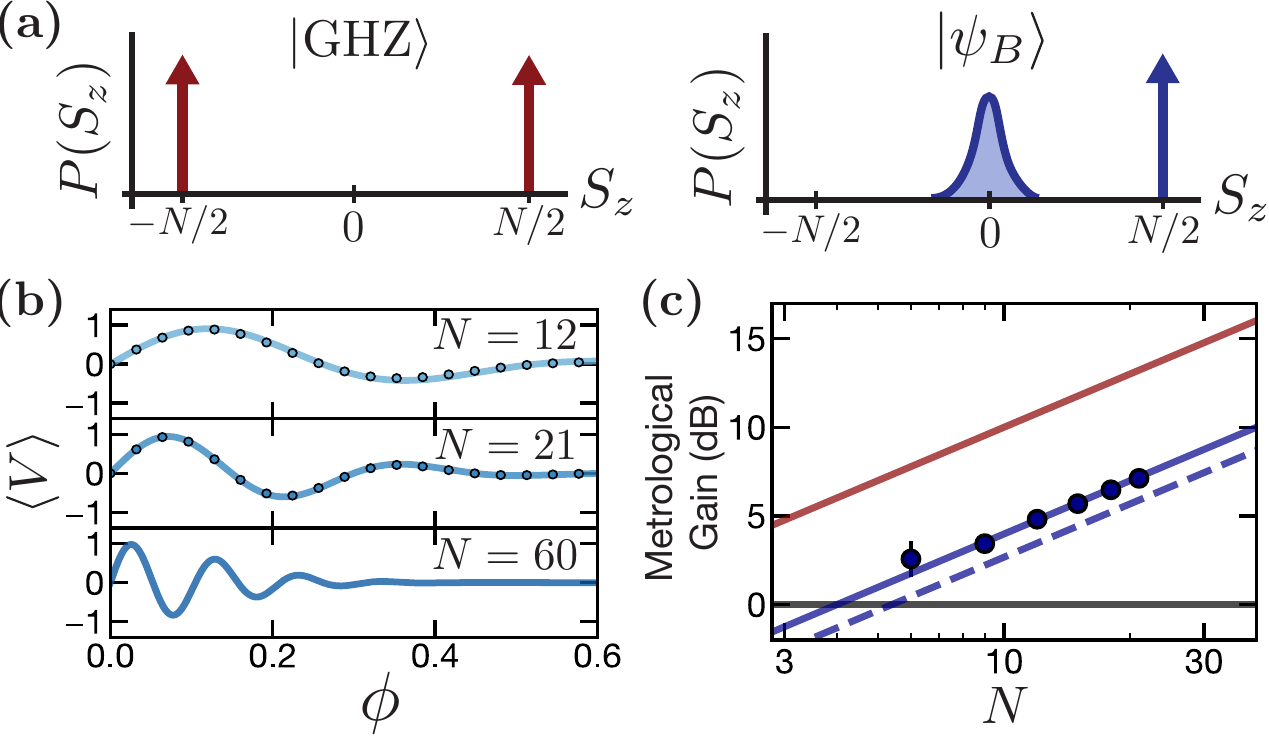}
 \caption{\textbf{(a)} Schematic of the polarization distribution, $P(S_z)$, of the GHZ state (left) and the butterfly state (right). 
 The former features two maximally separated peaks at $S_z = \pm N/2$; this leads to an optimal sensitivity, $\eta^{-1} = N$. 
 The latter features a narrow peak at $N/2$ and a broader peak at zero; this leads to a sensitivity, $\eta^{-1} \approx N/2$.
 \textbf{(b)} Numerical plot of the local expectation value, $\langle V \rangle_\phi$, as a function of the signal strength, $\phi$, when $U$ is Haar-random (solid lines) and when $U$ corresponds to evolution under a generic spin chain Hamiltonian for long times (dots)~\cite{supp}. The expectation value undergoes damped oscillations with frequency $\approx N/2$. \textbf{(c)} 
 The metrological gain, $1/(N \eta^2)$, of butterfly metrology with  local controls for a generic spin chain Hamiltonian as a function of system size (dots). The Heisenberg limit (red), standard quantum limit (black), and Haar-random predictions for local (solid blue) and global (dashed blue) 
 butterfly metrology  are shown for comparison. 
 }
\label{fig:haar}
\end{figure}

Our approach begins by introducing a new class of metrological state, 
which we call the `butterfly state':
\begin{equation} \label{eq:psi_B}
\ket{\psi_B} = \frac{\ket{\textbf{0}} + i {V}(t) \ket{\textbf{0}}}{\sqrt{2}}.
\end{equation}
Here, $\ket{\textbf{0}} \equiv \ket{0}^{\otimes N}$ is the fully polarized state, and $V(t) = U^\dagger  V U$ is a local Pauli operator (e.g.~$ V = \sigma_0^x$) evolved  under a unitary, $U$.  
We envision $U = e^{-iHt}$ to be generated by time evolution via a generic, many-body Hamiltonian, $H$.
For sufficiently late times, $U$ is  ``fully scrambling'' and resembles a Haar-random unitary. 
%
%, Floquet dynamics, or a digital quantum circuit.
% 
As depicted in Fig.~\ref{fig1}(a), $\ket{\psi_B}$ can be prepared by applying a local rotation $(\mathbbm{1}+i{V})/\sqrt{2}$ (i.e.~a $\pi/2$-pulse), sandwiched between forward and backward time evolution.

Physically, the butterfly state  is a superposition between two quantum trajectories generated by the two terms, $\mathbbm{1}$ and $i{V}$, of the local rotation [Fig.~\ref{fig1}(a)]. 
In the first, the forward and backward evolution perfectly cancel, returning the system to the  polarized state.
In the second, the local operator $i V$ disrupts this cancellation.
When $U$ is fully scrambling, this disruption will propagate to every spin in the system, and the second trajectory will resemble a Haar-random state, with an average polarization of zero. 
Thus, much like the GHZ state, the butterfly state $\ket{\psi_B}$ is a superposition of two states with a macroscopic difference, $\approx N/2$, in polarization [Fig.~\ref{fig:haar}(a)].
This yields a standard deviation, $\Delta S_z \approx N/4$, and thus an optimal sensitivity of $\eta \approx 2/N$, a mere factor of two from the Heisenberg limit.

\emph{Measurement protocol---}Having defined the butterfly state and its optimal sensitivity, we now devise an explicit protocol for measuring the external signal, $\phi$.
The simplest conceptual approach is to directly extract $\phi$ from a  measurement on the perturbed state.
However, this will require probing highly non-local coherences between the two trajectories in $\ket \psi_B$. 
This is even more difficult in our setting, compared to that of the GHZ state, since the second trajectory in $\ket \psi_B$ is a highly-entangled scrambled state.

To this end, we propose an alternative approach where we perform a final step of  time evolution under $U$ after the signal has been acquired [Fig.~\ref{fig1}(a)], and then conclude by measuring the \emph{local} operator $M \equiv V$~\footnote{We note that the quantum circuit implementing this approach [i.e.~Fig.~\ref{fig1}(a)] is closely related to the one-sided implementation of the teleportation circuit shown in Ref.~\cite{schuster2022many}}. 
This produces an expectation value,
\begin{align} \label{eq:V}
\langle V \rangle_\phi = &\frac 1 2 \bra{\textbf{0}} V(t) \ket{\textbf{0}} - \frac 1 2 \bra{\textbf{0}} V(t) e^{i \phi S_z} V(t) e^{-i \phi S_z} V(t) \ket{\textbf{0}} \nonumber \\ 
& + \textrm{Im} \left [ e^{i \phi N/2} \bra{\textbf{0}} V(t) e^{-i \phi S_z} V(t) \ket{\textbf{0}} \right ].
\end{align}
%where the latter expression holds to leading order in $\phi$.
For a small signal $\phi$, the first and second terms cancel, giving $\langle V \rangle_\phi \approx  \, \phi ( N/2 - \bra{\textbf{0}} V(t) S_z V(t) \ket{\textbf{0}} )$, and thus a sensitivity,
\begin{equation} \label{eq: sensitivity}
    \eta^{-1}_{\phi=0} =  N/2 - \bra{\textbf{0}} V(t) S_z V(t) \ket{\textbf{0}},
\end{equation}
which is precisely equal to the difference in polarization between the two trajectories in $\ket{\psi_B}$.
When the unitary is fully scrambling, the state ${V}(t) \ket{ \textbf{0}}$, has a mean polarization of zero, and our measurement protocol gives  $\eta_{\phi=0} \approx 2/N$; this  saturates the optimal sensitivity allowed by the butterfly state's QFI. 

\emph{Sensitivity from operator growth}---Thus far, we have focused on late times where the unitary $U$ is \emph{fully} scrambling. 
To characterize the sensitivity  at earlier times, we leverage a  precise connection between the sensitivity and information scrambling.
In particular, using $\sigma_i^z \ket{ \textbf{0} } = 1$, one can re-write
\begin{align} \label{eq:eta}
\eta^{-1}_{\phi=0} & = \frac 1 2  \sum_i \left( 1 - \bra{\textbf{0}} \sigma_i^z V(t) \sigma_i^z V(t) \ket{\textbf{0}}\right),
\end{align}
where the expectation values correspond to local out-of-time-order correlation functions (OTOCs)~\cite{kitaev_simple_2015,shenker2014black,roberts2015localized}.

These OTOCs quantify whether the time-evolved perturbation, $V(t)$, commutes with each spin operator $\sigma^z_i$.
As a function of time, each OTOC begins at unity and decays to zero as $V(t)$ grows to have support on spin $i$~\cite{landsman2019verified,blok2021quantum,schuster2022many}.
Thus, the sum in Eq.~(\ref{eq:eta}) counts the number of spins, $\mathcal{S}$, within the support of $V(t)$.
Intuitively, the operator $V(t)$ scrambles each spin within its support, effectively randomizing its polarization. 
This leads to a polarization difference $\sim \! \mathcal{S}/2$ between the scrambled state, $ V(t) \ket{\textbf{0}}$, and the fully polarized state, $\ket{\textbf{0}}$, yielding a sensitivity $\eta \approx 2 / \mathcal{S}$.

To this end, in order to evaluate the sensitivity of butterfly metrology at earlier times, one simply needs to understand the dynamics of operator growth under the Hamiltonian, $H$.
For Hamiltonians with local interactions, the operator support typically grows ballistically in time, $\mathcal{S} \approx (v_B t)^d$, where $d$ is the spatial dimension and $v_B$ is known as the butterfly velocity.
This growth continues until the so-called \emph{scrambling time}, $t_s \approx v_B^{-1} N^{1/d}$, after which the unitary is fully scrambling and the operator has support on the entire system, $\mathcal{S} \approx N$.
This leads to a sensitivity:    
\begin{equation}
    \eta \approx
    \begin{cases}
        2 / (v_B t)^{d}, & t \lesssim t_s \\
        2 / N, & t \gtrsim t_s, \\
    \end{cases}
    \label{eta_local}
\end{equation}
which continuously improves as a function of  time and saturates to its optimal value at $t_s$ [Fig.~\ref{fig:haar}(d)].

A few remarks are in order.
First, for Hamiltonians with long-range interactions, operators may exhibit faster-than-ballistic growth~\cite{zhou2023operator}.
For instance, under all-to-all-interactions, operators typically grow exponentially in time, $\mathcal{S} \sim e^{\lambda t}$, where $\lambda$ is the Lyapunov exponent~\cite{maldacena2016bound,sekino2008fast}.
This leads to a more rapid improvement in the sensitivity and an earlier saturation to its optimal  value.

Second, let us analyze the \emph{dynamic range} of butterfly metrology,~i.e.~its behavior at larger values of $\phi$.
As per Eq.~(\ref{eq:V}), this naturally requires us to consider the polarization distribution, $P(S_z)$, of the scrambled trajectory [Fig.~\ref{fig:haar}(a)].
For a Haar-random state, this polarization distribution is a Gaussian with mean zero and width $\sim \! \sqrt N$.
When the signal is small compared to this width, $\phi \ll 1/\sqrt{N}$, one can approximate $e^{i\phi S_z} V(t) \ket{\textbf{0}} \approx  V(t) \ket{\textbf{0}}$, leading to a sinusoidal expectation value $\langle V \rangle_\phi \approx \sin(\phi N/2)$ [Fig.~\ref{fig:haar}(b)], and a Heisenerg-scaling sensitivity.
For larger signal strengths, $\phi \gtrsim 1/\sqrt{N}$, this approximation breaks down and we find that the magnitude of the oscillations, and thus, the sensitivity, gradually decays to zero [Fig.~\ref{fig:haar}(b)]. 

Finally, the presence of conservation laws in the Hamiltonian will cause the sensitivity to deviate slightly from our estimates above. 
In particular, since time evolution cannot change the expectation value of any conserved quantity, the scrambled butterfly trajectory will  generally not approach a Haar-random state.
Instead, it will resemble a random state drawn from the Gibbs ensemble determined by the values of each conserved quantity in the initial state $\ket{\textbf{0}}$.
At late times, this still leads to a Heisenberg-scaling sensitivity, but with a pre-factor determined by the polarization density, $m$, of the Gibbs state, $\eta^{-1} \approx N(1-m)/2$.

\begin{figure}[t]
\includegraphics[width=\linewidth]{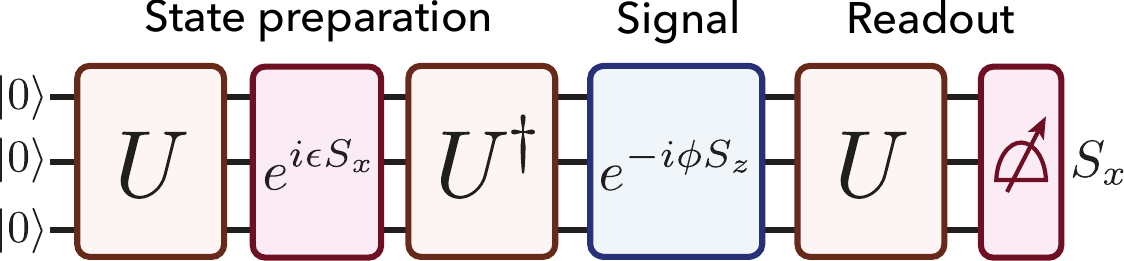}
\caption{Schematic of butterfly metrology with global controls. The metrological state is prepared using a global rotation, $e^{i\epsilon  S_x}$, under the spin operator $ S_x = \frac{1}{2} \sum_i \sigma^x_i$, and the signal is readout via the global expectation value, $\langle  S_x \rangle$.
}
\label{fig:global}
\end{figure}

\emph{Butterfly metrology with global controls---}Our discussions above have focused on the use of local controls within the butterfly metrology protocol: a local rotation to prepare the state, $|\psi_B\rangle$, and a local measurement to extract the signal, $\phi$.
We now demonstrate a variant of our protocol that utilizes purely global controls (Fig.~\ref{fig:global}); this is particularly important from an experimental perspective and allows butterfly metrology to be deployed in platforms without single-site addressing~\cite{hume2013accurate,davis2016approaching}.

In particular, instead of applying a local $\pi/2$-pulse to prepare the butterfly state (Fig.~\ref{fig1}), one applies the global rotation $e^{i \epsilon S_x}$, with $S_x = \frac 1 2 \sum_i \sigma_i^x$.
Similarly, in order to readout this signal, one measures the total polarization along the $x$-direction, $\langle {S}_x \rangle$.

To build some intuition for the sensitivity of our global butterfly metrology protocol, let us consider two limits.
First, consider the ``trivial'' limit where $t=0$ and $U = \mathbbm{1}$.
By setting $\epsilon = \pi/4$, one immediately recognizes that our protocol 
reduces to $N$ parallel copies of Ramsey spectroscopy, and thus exhibits an SQL-scaling sensitivity, $\eta \sim 1/\sqrt{N}$.

Second, consider the late-time limit where $U$ is fully scrambling.
Much as before, we decompose the global rotation into an identity and non-identity component, $e^{i \epsilon S_x} \equiv a \mathbbm{1} +i \tilde{V}$, where $a = \cos^N(\epsilon)$ and $\tilde V$ is traceless. 
This leads to the ``global'' butterfly state,
\begin{equation}
| \tilde \psi_B \rangle = a \ket{\textbf{0}} +i \tilde{V}(t) \ket{\textbf{0}}.
\end{equation}
In order to maximize the sensitivity of our global protocol, we set $\epsilon \sim 1/\sqrt{N}$ so that the butterfly state, $| \tilde \psi_B \rangle$, is an approximately equal superposition of its two trajectories.
More specifically, we find that the optimal value occurs at $\epsilon = 1/\sqrt{2N}$, which yields a  Heisenberg-scaling sensitivity, $\eta = \sqrt{2e} / N \approx 2.3/ N$~\cite{supp}.
We note that this is nearly identical to the late-time sensitivity obtained in our local protocol [Eq.~(\ref{eta_local})]. 

As one evolves from time $t=0$ to the scrambling time $t= t_s$, the sensitivity smoothly interpolates between the two limits above. 
This interpolation can again be understood in the language of operator growth; indeed, at intermediate times, the sensitivity is given by $\eta \sim 1/\sqrt{N \mathcal{S}}$, where $\mathcal{S}$ is the number of spins in the  support of a time-evolved local operator~\cite{supp}. 

\begin{figure}[t]
\includegraphics[width=\linewidth]{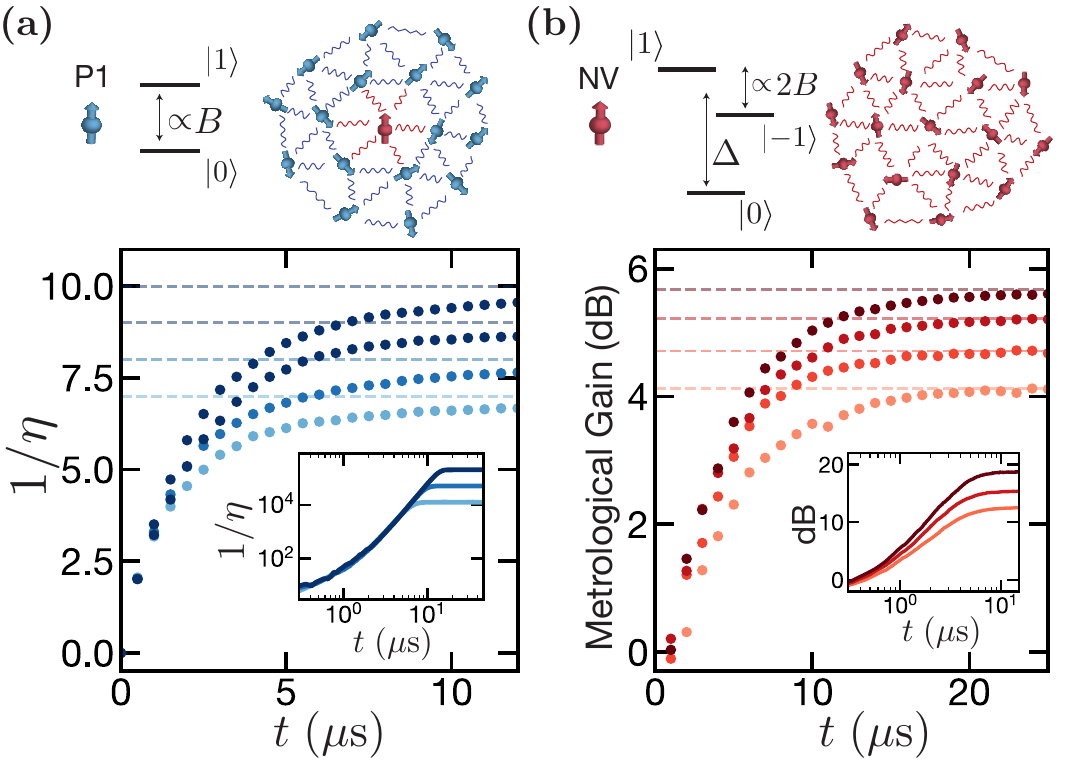}
\caption{Numerical simulations for the sensing protocol under the dynamics of two experimental platforms. \textbf{(a)}  The sensitivity of our protocol with \emph{local} controls for a hybrid spin system, consisting of a single NV center surrounded by a cluster of P1 centers. The simulations are performed via exact diagonalization with $N \in [14,20]$ total spins. After an initial growth period, the sensitivity saturates at $\eta = 2/N$ (dashed line), consistent with our expectation for fully scrambled dynamics. The inset displays the sensitivity for large-scale systems, $N \sim 10^4-10^5$, simulated via a stochastic growth model~\cite{supp}. \textbf{(b)} The metrological gain, $1/(N \eta^2)$, and sensitivity (inset) of our protocol with \emph{global} controls implemented for a dense ensemble of NV centers. The total number of spins is $N \in [14,20]$ spins. For both systems, the density of spin defects is 100 ppm, corresponding to an average separation of $\sim 4$ nm.  
}
\label{fig:expt}
\end{figure}

\emph{Experimental proposals---}Perhaps the most unique feature of butterfly metrology is its universality---the protocol utilizes evolution under a  generic many-body Hamiltonian and thus, can naturally be realized in a wide variety of experimental platforms (Table~\ref{tab:expts}). 
Here, we focus on the setting of solid-state spin ensembles~\cite{doherty2013nitrogen}, where achieving quantum-enhanced sensitivities remains an outstanding challenge~\cite{goldstein2011environment,cooper2019environment,degen2021entanglement}. 
A key part of this challenge stems from the nature of the dipolar interaction intrinsic to such systems.
Indeed, for disordered spin ensembles, the dipolar Hamiltonian, while strongly-interacting, is far from the conventional all-to-all-coupled models (e.g.~one-axis twisting) typically used to generate metrologically-useful entanglement~\cite{davis2016approaching,colombo2022time}.

To this end, we propose and analyze two implementations of butterfly metrology using dense ensembles of nitrogen-vacancy (NV) color centers in diamond~\cite{schirhagl2014nitrogen,doherty2013nitrogen}.
NV centers behave as sensitive magnetometers and have been used to detect a variety of external signals, ranging from neuronal action potentials to superconducting vortices~\cite{barry2016optical,schlussel2018wide}. 
Each NV exhibits a spin, $S=1$, electronic ground state [Fig.~\ref{fig:expt}(b)], which can be optically polarized and read out at room temperature. 
In the absence of an external magnetic field, the NV's spin sublevels are separated by $\Delta = 2.87$~GHz, corresponding to the zero-field splitting~\cite{doherty2013nitrogen}.

In addition to NV centers, the diamond sample also contains a relatively high concentration of optically-dark, spin-1/2 P1 centers (substitutional nitrogen impurities). 
Indeed, each NV center is typically surrounded by a large bath of P1 spins [Fig.~\ref{fig:expt}(a)]~\cite{hall2016detection,zu2021emergent}. 
Conventionally, such P1s are viewed as a source of decoherence and dynamical decoupling is employed to suppress their interactions with NVs~\cite{choi2020robust}.

For our first implementation, we describe a realization of butterfly metrology (with local controls) in a hybrid NV-P1 system, where the NV center provides the local control and the P1 centers comprise the remaining spins~[Fig.~\ref{fig:expt}(a)]. 
We envision an experimental scenario where the P1 centers are polarized either by working at cryogenic temperatures or via Hartmann-Hahn hyperpolarization~\cite{belthangady2013dressed}.~For external magnetic fields, $B$, away from level anti-crossing, the NV center is off-resonant from the P1s, enabling the application of  selective local rotations via microwave fields.
This off-resonance also affects the many-body interactions in the system: the P1s interact with one another via the full magnetic dipolar interaction, and with the NV, via the Ising term of this interaction~\cite{supp}. 
Crucially, both of these  interactions can be time reversed using a simple generalization of the so-called WAHUHA pulse sequence~\cite{supp}.

For our second implementation, we consider a scenario where the NV-P1 interactions have been dynamically decoupled, leaving only interactions \emph{between} NV centers themselves~\cite{kucsko2018critical,de2010universal}. 
This scenario is naturally suited to the global-variant of butterfly metrology [Fig.~\ref{fig:global}].
In particular, we envision applying a global rotation on the entire ensemble of NV centers using a microwave pulse. 
As before, time-reversed dynamics can be achieved via Hamiltonian engineering techniques~\cite{supp}. 

\begin{table}
\centering
\begin{tabular}{ l | c |  c}
\hline
Experimental platform  & Control & Time-reversal method \\
\hline
\hline
Hybrid NV-P1 system~\cite{zu2021emergent} & Local  & Hamiltonian eng. \\
NV-center ensemble~\cite{kucsko2018critical}  & Global  & Hamiltonian eng. \\
Dipolar Rydberg atoms~\cite{chen2023continuous} \! & Both  & State encoding\\
Atoms in optical cavity~\cite{periwal2021programmable}\! &  Global  & Laser detuning \\
Superconducting qubits~\cite{braumuller2022probing}\! & Both &  $\pi$-pulse \\
Trapped ions~\cite{pogorelov2021compact,egan2021fault}  & Both & Laser phase \\
\hline
%\hline
\end{tabular}
\caption{Butterfly metrology with either local or global controls can be realized in a wide array of experimental platforms. Detailed blueprints and discussions are provided in the supplemental materials~\cite{supp}.
}
\label{tab:expts}
\end{table}

To explore the dynamics of the sensitivity in both implementations, we perform an extensive set of numerical simulations using both Krylov subspace methods (for smaller system sizes) and a stochastic model for operator growth (for larger system sizes)~\cite{chen2019quantum,zhou2020operator,zhou2023operator}. 
For our hybrid NV-P1 approach [Fig.~\ref{fig:expt}(a)], the sensitivity improves rapidly in time and  saturates at $\eta \approx 2/N$, consistent with our previous analysis. 
The sensitivity dynamics of our global NV-based approach [Fig.~\ref{fig:expt}(b)] are somewhat similar.
In this case, we plot the  metrological gain, $G = \eta^{-2}/N$, which saturates at  $G \approx (0.43)^2 N$ and quantifies the enhancement compared to the standard quantum limit~\footnote
{For the global protocol, we plot the metrological gain, instead of the sensitivity, since the global protocol has a sensitivity given by the SQL at time zero.}.
Interestingly, for larger system sizes (up to $\sim10^6$ spins), we find that both the sensitivity (local variant) and  metrological gain (global variant) exhibit a sustained period of super-polynomial growth.
This arises from the Levy flight growth of operators in 3D with $1/r^3$ interactions~\cite{zhou2023operator}.

Much like other time-reversal-based sensing protocols~\cite{goldstein2011environment,davis2016approaching}, butterfly metrology maintain a Heisenberg-limited sensitivity in the presence of both read-out and initialization errors. 
Our protocol is particularly robust against coherent errors that can be time reversed; for example, slow variations in the Hamiltonian parameters~\cite{pogorelov2021compact}.
The effect of incoherent errors can be understood using our framework of operator growth, namely, they suppress the sensitivity by a factor $\sim e^{-4 \gamma \int_0^t dt' \mathcal{S}(t')}$~\footnote{We note that this suppression factor is precisely the square of the so-called Loschmidt echo~\cite{schuster2022operator}.}, where $\gamma$ is the local noise rate and $\mathcal{S}(t)$ is the operator support~\cite{schuster2022operator,supp}.

Our work opens the door to a number of intriguing future directions.
First, since it is only limited by the operator light-cone, butterfly metrology represents the fastest way to generate metrologically-useful entanglement in any given  interaction geometry~\cite{supp}.
For example, compared to recent work leveraging continuous symmetry breaking in two-dimensional dipoles~\cite{block2023universal,chen2023continuous}, butterfly metrology offers a quadratic speed-up to achieve the same enhanced sensitivity. 
Second, we note that achieving any sensitivity scaling beyond the standard quantum limit guarantees multi-partite entanglement~\cite{hyllus2012fisher}.
Thus, from the perspective of entanglement verification~\cite{guhne2009entanglement}, 
butterfly metrology provides a generic strategy for witnessing  the entanglement generated by many-body quench dynamics. 
Finally, although we have focused on analog Hamiltonian dynamics, we note that butterfly metrology can also be realized using digital quantum circuits.
In this setting, our protocol's robustness to coherent errors could be particularly advantageous for mitigating the effects of low-frequency noise and imperfect gate calibration~\cite{pogorelov2021compact,krinner2020benchmarking}.

\textit{Acknowledgements}---We gratefully acknowledge the insights of and discussions with Monika Schleier-Smith, Dan Stamper-Kurn, Francisco Machado, Antoine Browaeys, Marcus Bintz, Bingtian Ye, and Zilin Wang.
This work was supported in part by multiple grants from the ARO  
(MURI program grant no.~W911NF-20-1-0136 and grant no.~W911NF-21-1-0262), and the Brown Institute for Basic Sciences. 
T.S. acknowledges
support from the Walter Burke Institute for Theoretical Physics at Caltech.
N.Y.Y. acknowledges support from a Simons Investigator award.

\bibliography{refs}

\begin{thebibliography}{84}
\bibitem [{spi()}]{spin_basis}%
  \BibitemOpen
  \href@noop {} {\bibinfo  {journal} {For spin-$1/2$ particles, we adopt the
  convention that $\ket{0}$ is the ``spin up'' state, and $\ket{1}$ is the
  ``spin down'' state; i.e., $\sigma_z \ket{0} = \ket{0}$ and $\sigma_z \ket{1}
  = - \ket{1}$}\ }\BibitemShut {NoStop}%
\end{thebibliography}

\end{document}

\title{Supplemental Material: A Universal Protocol for Quantum-Enhanced Sensing via Information Scrambling}

\author{Bryce Kobrin}
\thanks{These authors contributed equally to this work.}
\affiliation{Department of Physics, University of California Berkeley, Berkeley, CA 94720, USA}

\author{Thomas Schuster}
\thanks{These authors contributed equally to this work.}
\affiliation{Walter Burke Institute for Theoretical Physics and Institute for Quantum Information and Matter, California Institute of Technology, Pasadena, CA 91125 USA}
\affiliation{Department of Physics, University of California Berkeley, Berkeley, CA 94720, USA}

\author{Maxwell Block}
\affiliation{Department of Physics, Harvard University, Cambridge, MA 02138, USA}

\author{Weijie Wu}
\affiliation{Department of Physics, Harvard University, Cambridge, MA 02138, USA}

\author{Bradley Mitchell}
\affiliation{IBM Quantum, IBM Almaden Research Center, San Jose, CA 95120, USA}

\author{Emily Davis}
\affiliation{Department of Physics, Harvard University, Cambridge, MA 02138, USA}
\affiliation{Department of Physics,
New York University, New York, NY, 10003, USA}

\author{Norman Y.~Yao}
\affiliation{Department of Physics, Harvard University, Cambridge, MA 02138, USA}

\date{\today}

\maketitle

\maketitle
\tableofcontents

\section{Comparison to existing time-reversed sensing protocols}

\begin{figure}[t]
\includegraphics[width=0.5\linewidth]{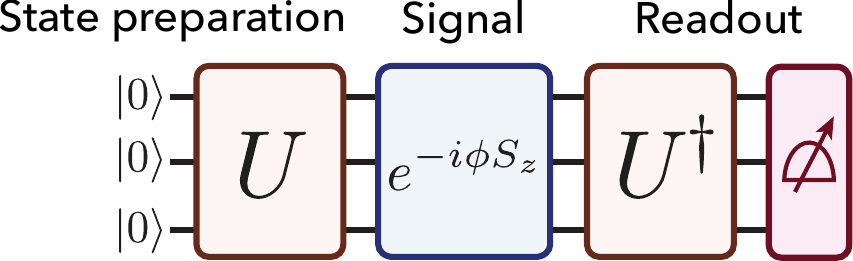}
\caption{Schematic depiction  of existing time-reversal-based protocols for quantum-enhanced sensing~\cite{davis2016approaching,macri2016loschmidt,mooney2021generation,li2023improving}. Forward time evolution is used to prepare the metrological state, and reverse evolution is applied solely during readout.}
\label{fig:echo}
\end{figure}

Many of the most prominent protocols for quantum-enhanced sensing involve the use of time-reversed dynamics, through variants of the so-called echo protocol~\cite{davis2016approaching,macri2016loschmidt,mooney2021generation,li2023improving} (depicted in Fig.~\ref{fig:echo}).
These protocols all feature a crucial difference from our proposed butterfly metrology protocol, in that they utilize only a \emph{single step each} of forward and reverse time evolution.
Namely, the metrological state is prepared via forward evolution from a product state, $U\ket{\textbf{0}}$, and reverse evolution is applied solely to assist readout.
As a result, such protocols can only achieve a quantum enhancement for extremely specific classes of unitary dynamics, i.e.~those that generate a large quantum Fisher information (QFI) when applied to a product state. 
In contrast, butterfly metrology utilizes a combination of forward and reverse time evolution to \emph{prepare} a metrologically-useful state; as we show in our work, this enables a Heisenberg-scaling quantum enhancement for completely generic interacting dynamics.

To illustrate this distinction, let us recall a few classes of unitaries $U$ for which the standard echo protocol has previously been applied.
First, take $U$ to be a Clifford circuit that prepares a GHZ state, i.e.~$U \ket{\textbf{0}} =  (\ket{0}^{\otimes N} + \ket{1}^{\otimes N})/\sqrt{2}$.
The full sensing scheme consists of applying $U$ to generate a GHZ state, accumulating a phase under the external signal, and then applying the inverse preparation circuit to refocus the acquired phase to a single-body observable. 
This last step is not strictly necessary---one could instead readout the signal via global parity measurements of the GHZ state; however, readout via a local observable provides a practical advantage as it leads to much greater robustness to readout noise\footnote{Interestingly, sensing with a GHZ state can be understood as a special case of our protocol. Consider a protocol which prepares the GHZ state by applying a $\pi/2$-pulse to the first qubit, $( \mathbbm{1} + i \sigma_0^x)/\sqrt{2}$, followed by a CNOT ``ladder'' denoted $U$. To detect the accumulated phase from an external signal, the inverse ladder $U^\dagger$ is applied and the first qubit is measured. The final state (up to normalization) is $U^\dagger e^{i \phi S_z} U ( \mathbbm{1} + i \sigma_0^i) \ket{\textbf{0}}$. Because $\ket{\textbf{0}}$ is an eigenstate of the CNOT gates, we can insert an additional copy of $U^\dagger$ at the beginning of the circuit without changing the final outcome: $U^\dagger e^{i \phi S_z} U ( \mathbbm{1} + i  \sigma_0^i) U^\dagger \ket{\textbf{0}}$. This is precisely of the form of our protocol, upon swapping $U \leftrightarrow U^\dagger$.}\cite{mooney2021generation}.
Second, take $U= e^{-iHt}$ to be time evolution under a large-spin Hamiltonian, e.g.~the one-axis twisting model, $H=S_z^2$ with $S_z = \frac 1 2 \sum_i \sigma_i^z$. 
Such Hamiltonians are governed by effectively semi-classical dynamics owing to the large-spin degree of freedom.
These dynamics can be used to generate metrological spin-squeezed states from solely forward time evolution~\cite{davis2016approaching,li2023improving}.
Much like for the GHZ state, the effect of the signal can in principle be detected directly, via measurements on the squeezed state; however, in practice, applying the inverse preparation, $U^\dagger = e^{iHt}$, to ``un-squeeze'' the state substantially improves the robustness to readout noise~\cite{davis2016approaching,colombo2022time}.
Additional examples of unitary dynamics proposed for the echo protocol include time evolution under certain integrable Hamiltonians~\cite{goldstein2011environment} or Hamiltonians that feature a continuous-symmetry-breaking phase~\cite{block2023universal}.

Crucially, however, the echo protocol shown in Fig.~\ref{fig:echo} \emph{does not} lead to any metrological enhancement for generic interacting quantum dynamics, outside of these few, highly specific cases.
To illustrate this limitation in a simple example, consider when $U$ is taken to be a Haar-random unitary.
This serves as a standard model for the late-time dynamics of generic interacting quantum systems. 
In this case, the state prepared under forward evolution, $U \ket{\textbf{0}}$ is a Haar-random state.
A Haar-random state has an average Fisher information
\begin{equation}
\begin{aligned}
    \mathbbm{E}_\psi \left[ \mathcal{F} \right] &= 2 \mathbbm{E}_\psi \left[ \bra{\psi} S_z^2 \ket{\psi} - \bra{\psi} S_z \ket{\psi}^2 \right] \\
    & = \frac{1}{2^N} \tr( S_z^2 ) - \frac{1}{2^N(2^N+1)} \left( \tr( S_z )^2 + \tr( S_z^2 ) \right) \\
    & = \frac{N}{4} + \mathcal{O}\left( \frac{N}{2^N} \right),
\end{aligned}
\end{equation}
which implies a sensitivity that is bounded by the standard quantum limit.
Intuitively, this follows because a random state, with high probability, has no long-range correlations in the $z$-basis.
%
%This implies that the QFI scales as $\mathcal{F} \sim N$, and thus a sensitivity that is bounded by the standard quantum limit.

As discussed in the main text and shown in detail below, our protocol circumvents this restriction, enabling Heisenberg-scaling sensitivities for any generic interactings Hamiltonian dynamics. 
This greater versatility opens the door to achieving a metrological enhancement in a much wider variety of experimental platforms (e.g.~the spin systems discussed in Sections \ref{sec:NV-P1} and \ref{sec:NV}).
Furthermore, even in systems that can achieve a metrological enhancement using existing echo protocols, our approach can offer several advantages with respect to the preparation time, and robustness to coherent errors, as discussed further in Sections \ref{sec:SC} and \ref{sec:traped-ion}.

\section{Detailed analysis of the sensitivity}

Here, we provide further details on our calculations of the sensitivity of both the local and global variants of butterfly metrology.

\subsection{Butterfly metrology with local control}

We begin with the local protocol.
As discussed in the main text, the signal of the local protocol takes the following form,
\begin{equation} \label{eq:V}
\begin{split}
\langle V \rangle_\phi = &\frac 1 2 \bra{\textbf{0}} V(t) \ket{\textbf{0}} - \frac 1 2 \bra{\textbf{0}} V(t) e^{i \phi S_z} V(t) e^{-i \phi S_z} V(t) \ket{\textbf{0}} \\ 
& + \textrm{Im} \left [ e^{i \phi N/2} \bra{\textbf{0}} V(t) e^{-i \phi S_z} V(t) \ket{\textbf{0}} \right ].
\end{split}
\end{equation}
where $V$ is a Pauli operator and $V(t) = U^\dagger V U$.
At late times, we approximate the evolution $e^{-iHt}$ by a Haar-random unitary $U$.
For such a unitary, the first two terms in the signal vanish in expectation, leaving only the final term non-zero. 
To analyze this term, we decomposed the perturbed state in the computational basis as $ V(t) \ket{\textbf{0}} = \sum_{s \in \{0,1\}^N} c_s \ket{s}$, and define the \emph{polarization distribution} $P(S_z) = \sum_{|s|=2 S_z} \left|c_s \right|^2$, where $|s|= \sum_i (-1)^{s_i}$ and  $S_z = -N/2,-N/2+1,\ldots,N/2-1,N/2$.
We observe that the signal is determined by the characteristic function of the distribution,
\begin{align} \label{eq:im_Phi}
\langle V \rangle_\phi &= 
\textrm{Im} 
\big [ e^{i \phi N/2} 
\sum_{S_z=-N/2}^{N/2} e^{-i \phi S_z} P(S_z) \big ] \\ 
&= \textrm{Im} 
\big [ e^{i \phi N/2} \Phi(\phi) \big ]
\end{align}
where $\Phi(\phi) = \sum_{S_z=-N/2}^{N/2} e^{-i \phi S_z} P(S_z )$. 
Since $U$ is a Haar-random unitary, we can approximate the perturbed state as a Haar-random state.
This leads to a binomial polarization distribution, $P(S_z) = \frac 1 {2^{N}} {N \choose N/2-S_z}$. 
Plugging this distribution into Eq.~(\ref{eq:im_Phi}) allows us to compute the expected signal as a function of $\phi$, as shown in Fig.~2(b) of the main text.

To analyze the sensitivity for small values of $\phi$, we Taylor expand Eq.~(\ref{eq:im_Phi}) to leading order. The sensitivity is determined by the first moment of $P(S_z)$,
\begin{equation} \label{eq:avg_size}
\eta^{-1}_{\phi=0} = N/2 -\sum_{S_z} S_z P(S_z).
\end{equation}
The polarization distribution for a Haar-random unitary has mean zero and thus leads to $\eta_{\phi=0} = 2/N$, a factor of 2 above the strict Heisenberg limit.

For larger $\phi$, two effects that cause the sensitivity to deviate from this maximal value.
The first effect is not particularly important, and arises simply because our signal, $\langle V \rangle_\phi$, is oscillatory as a function of $\phi$.
This causes the sensitivity to similarly oscillate between zero and its maximal value.
If desired, this effect can easily be mitigated by measuring the opposite quadrature of the oscillation (i.e.~measuring the real part of the characteristic function, $\textrm{Re} 
\left [ e^{i \phi N/2} \Phi(\phi) \right ]$, in addition to the imaginary part); we provide an explicit protocol to do so in Section~\ref{sec: opposite quadrature}.
The optimal sensitivity is obtained by taking a linear combination of the real and imaginary parts, $C(\phi,\theta) = \cos(\theta) \textrm{Re} 
\left [ e^{i \phi N/2} \Phi(\phi) \right ] + \sin(\theta) \textrm{Im} 
\left [ e^{i \phi N/2} \Phi(\phi) \right ]$, and maximizing $| \partial_\phi C(\phi,\theta) |$ with respect to $\theta$.
This yields an optimal sensitivity $\eta^{-1}_{\phi} = |-iN/2\Phi(\phi) + \Phi^\prime(\phi)|$.

The second effect of a larger $\phi$ is more fundamental, and arises due to the finite width of the polarization distribution, $P(S_z)$, about its mean value [see e.g.~Fig.~2(a) of the main text].
Intuitively, this non-zero width causes the magnitude of the expectation value, $\bra{\textbf{0}} V(t) e^{-i \phi S_z} {V}(t) \ket{\textbf{0}}$, to decrease below unity, since the state, ${V}(t) \ket{\textbf{0}}$, is no longer a perfect eigenstate of $e^{-i \phi S_z}$.
This damps the sensitivity by a factor of, roughly,
\begin{equation}
    \left| \bra{\textbf{0}}  V(t) e^{-i \phi S_z} {V}(t) \ket{\textbf{0}} \right| \approx 1 - \frac{1}{2} \phi^2 \left( \bra{\textbf{0}} V(t) S_z^2 {V}(t) \ket{\textbf{0}} - \bra{\textbf{0}} V(t) S_z {V}(t) \ket{\textbf{0}}^2 \right)
\end{equation}
where on the RHS we Taylor expand to second order in $\phi$.
The damping is controlled by $\phi^2$ multiplied by the variance of the polarization distribution.
In a Haar-random state, the polarization distribution has variance $\sim \! N$, which leads the sensitivity to decay from its maximal value for $\phi \gtrsim 1/ \sqrt{N}$.
In Fig.~2 of the main text, we corroborate this prediction by plotting the exact sensitivity with respect to $\phi$ for a Haar-random unitary.
As expected, we observe that the range for high sensitivity is $\phi \lesssim 1/\sqrt{N}$.
In the limit $N \gg 1$, we can compute the full functional form of the sensitivity by approximating the polarization distribution as a Gaussian, $P(S_z) \approx e^{-2S_z^2/N}$, which yields a sensitivity $\eta^{-1}_{\phi} \approx (N/2) e^{-\phi^2 N/8}$.

\begin{figure}[t]
\includegraphics[width=0.4\linewidth]{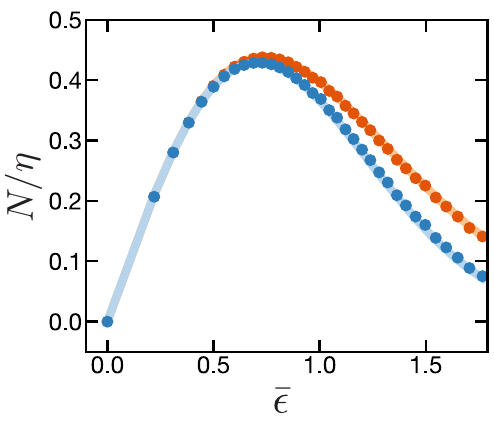}
\caption{The sensitivity of butterfly metrology with global controls as a function of the re-scaled preparation angle, $\bar \epsilon = \sqrt{N} \epsilon/2$, for the measurement operator $S_x$ (blue), and the measurement operator $\sin (\epsilon S_x)$ (orange). We observe excellent agreement between the analytic prediction for Haar-random evolution (solid line) and the late-time dynamics of an interacting Hamiltonian on $N=18$ spins (points). In this plot, the spin Hamiltonian consists of all-to-all two-body interactions with random magnitudes, $H=\sum_{i < j} \sum_{\mu,\nu} J_{ij}^{\mu \nu} \sigma^i_\mu \sigma^j_\nu$, where $\mu,\nu \in \{X,Y,Z\}$ and $J_{ij}^{\mu \nu} $ are Gaussian random numbers with mean zero and standard deviation $J/\sqrt{N}$. The evolution time is $tJ = 10$. Similar results are obtained for generic Hamiltonians, including translation-invariant Hamiltonians in 1D and 2D, although the agreement between such Hamiltonians and the Haar-random prediction is not as precise as above at the accessible system sizes.}
\label{fig:global_sensitivity}
\end{figure}

\subsection{Butterfly metrology with global control}

We now turn to the global variant of butterfly metrology.
The signal takes the form
\begin{equation}  \label{eq: signal global}
\begin{split}
\langle  S_x \rangle_\phi =& a^2 \bra{\textbf{0}}  S_x(t) \ket{\textbf{0}} - \bra{\textbf{0}} \tilde V(t) e^{i \phi S_z} S_x(t) e^{-i \phi S_z}  V(t) \ket{\textbf{0}} \\ 
& + 2 a \textrm{Im} \left [ e^{i \phi \frac N 2} \bra{\textbf{0}} S_x(t) e^{-i \phi S_z} \tilde V(t) \ket{\textbf{0}} \right ],
\end{split}
\end{equation}
where we decompose the global rotation as $e^{i \epsilon  S_x} \equiv a \mathbbm{1} +i \tilde{V}$, where $a = \cos^N(\epsilon)$ and $\tilde V$ is traceless, as described in the main text. 
We note two key differences between the signal, Eq.~(\ref{eq: signal global}), of the global protocol, and the signal, Eq.~(\ref{eq:V}), of the local protocol.
First, the global protocol contains the free parameter, $a$, which is set by our choice of the rotation angle, $\epsilon$.
Second, the third term in the signal of the global protocol involves a matrix element between two \emph{distinct} quantum states, $\tilde V(t) \ket{\textbf{0}}$ and $S_x(t) \ket{\textbf{0}}$.

Let us first address the behavior of the global protocol for late time, i.e.~Haar-random evolution, and then turn to earlier times.
Similar to the local protocol, when $U$ is Haar-random only the third term in the signal is non-vanishing. 
Moreover, within the third term, only the component of $\tilde V(t) \ket{\textbf{0}}$ that overlaps with $S_x(t) \ket{\textbf{0}}$ contributes to the matrix element.
In particular, if we decompose the global rotation as  $e^{i\epsilon S_x}= a \mathbbm{1} + i 2 a \tan(\epsilon/2) S_x + S_{x,\perp}$, where $\tr(S_x S_{x,\perp}) = 0$, then the Haar average retains only the term proportional to $S_x$.
Thus, the signal simplifies to 
\begin{equation} \label{eq:global}
\langle  S_x \rangle_\phi = 4 a^2 \tan(\epsilon/2) \cdot \textrm{Im} \big [ e^{-i \phi \frac N 2} \bra{\textbf{0}} S_x(t) e^{i \phi S_z} S_x(t) \ket{\textbf{0}} \big ].
\end{equation}
We note that, crucially, the state $S_x(t) \ket{\textbf{0}}$  has normalization $\bra{\textbf{0}} S_x(t)S_x(t) \ket{\textbf{0}} = N/4$.
% 
%Thus, in combination, the signal of the global protocol differs from local control by a factor of $a^2 N \tan(\epsilon/2)$.

Signal in hand, we now turn to the sensitivity, $\eta^{-1}_{\phi=0} \equiv (\partial_\phi \langle S \rangle_\phi/\Delta S_\phi)_{\phi = 0}$, of the global protocol.
For convenience, we work in the limit $N \gg 1$, such that $a = \cos^N(\epsilon/2) \approx \textrm{exp}(-\bar \epsilon^2/2)$, where $\epsilon \equiv 2 \bar \epsilon /\sqrt{N}$ and $\bar \epsilon$ is an order-one constant.
For small values of $\phi$, we have $\partial_\phi \langle S_x \rangle_{\phi=0} \approx a^2 \bar \epsilon N^{3/2}/2$.
Meanwhile, the standard deviation of the global measurement is $\Delta S_{\phi = 0} = \sqrt{N}/2$.
This yields $\eta^{-1}_{\phi=0} \approx \bar \epsilon e^{-\bar \epsilon^2} N$, with an optimal value $\eta^{-1} = (1/\sqrt{2e}) N \approx 0.43 N$ achieved at $\bar \epsilon = 1/\sqrt{2}$.
In Fig.~4(b) of the main text, we compare this prediction to the sensitivities obtained in exact numerical simulations of a spin model with $N=18$ spins and observe excellent agreement.

We remark that, interestingly, measuring the global spin operator, $S_x$, is in some cases not the optimal measurement for the global protocol.
In particular, for Haar-random evolution, and a very small improvement in sensitivity may be achieved by instead computing the expectation value of the operator, $M = \tilde V + \tilde V^\dagger = 2 \sin (\epsilon S_x)$.
Deriving the sensitivity with respect to a measurement of $M$ is straightforward but tedious, so we simply quote the result: $\eta^{-1}_{\phi=0} = (a/ \sqrt{2}) (1-a^4)^{1/2} N$, which yields an optimal sensitivity $\eta^{-1} = (1/3^{3/4}) N \approx 0.44 N$.
Although this offers a minute advantage over the original protocol, it may be more challenging to realize experimentally since it involves measuring higher powers of $S_x$. 

To understand the sensitivity of the global protocol at earlier times, we leverage the connection between butterfly metrology and operator growth introduced in detail in the main text.
To begin our analysis, we note that the global rotation $e^{i \epsilon  S_x}$ flips each spin in the system with probability $\sin^2(\epsilon/2) \sim \epsilon^2$.
Thus, the perturbation $\tilde{V}$ consists of $\sim \! \epsilon^2 N$ local spin operators at time zero, separated in space by a typical distance  $\sim \! 1/\epsilon^2$ (working in 1D for simplicity).
%
%This can be seen by expanding the global rotation as a Taylor series in $\epsilon$, and noting that the $\mathcal{O}(\epsilon N)$ terms dominate.
%
As we time evolve, each local operator grows to have support on a larger region of $\mathcal{S}$ spins.
For a given time, the sensitivity is optimized when we set $\epsilon \sim 1/\sqrt{\mathcal{S}}$, just small enough so that the operators remain separated after time evolution.
In this regime, the global protocol factorizes, to good approximation, into $N/\mathcal{S}$ parallel protocols on $\mathcal{S}$ spins each.
%so that the full time-evolved perturbation, $\tilde{V}(t)$, has support on $\epsilon^2 N \cdot \mathcal{S} \sim N$ spins.
%
Now, suppose that we set the evolution time (or conversely, set $\epsilon$) so that each local operator time evolves to have support on a region of size $\mathcal{S} \sim 1/\epsilon$.
In this case, the typical behavior of the global protocol resembles that of $\sim \! N/\mathcal{S}$ copies of the \emph{local} rotation protocol performed in parallel.
Each ``local'' protocol is evolved to the scrambling time on $\mathcal{S}$ spins, and thus has a Heisenberg-scaling sensitivity $\eta^{-1} \sim \! \mathcal{S}$.
These add in quadrature to give a total sensitivity $\eta^{-1} \sim \sqrt{N \mathcal{S}}$.
As we increase time, the size $\mathcal{S}$ smoothly increases from $1$ to $N$, yielding a sensitivity that smoothly improves from the SQL to the Heisenberg limit.
The precise functional form of this interpolation will depend on the growth of the operator support $\mathcal{S}$ under the specific Hamiltonian dynamics of interest; we refer to the main text for further discussion.

\begin{table}
\centering
\begin{tabular}{l | l c c c l}
\hline
\hline
Experimental platform  & Form of interaction & Geometry & Control & Initial state & Mechanism for time reversal \\
\hline
Dipolar Rydberg atoms & Dipolar XY & 1D or 2D & Either & Random & Rydberg-state encoding\\
Hybrid spin system & Dipolar Ising/XXZ & 3D with disorder & Local & Polarized & Hamiltonian engineering \\
Ensemble of NV centers & Dipolar XXZ & 3D with disorder & Global & Polarized & Hamiltonian engineering \\
Atoms in optical cavity & Long-range XY & Variable & Global & Polarized & Sign of laser detuning \\
Superconducting quits & Local XY & 2D & Either & Random & Conjugation by pi-pulses \\
Trapped ions & Digital gates & Variable & Either & Either & Phase of laser excitation \\
\hline
\hline
\end{tabular}
\caption{A more detailed overview of our proposed implementations of butterfly metrology in variety of experimental platforms. The range of interaction forms and system geometries demonstrates the versatility of butterfly metrology. For certain platforms, we observe that it is helpful to randomize the initial state in a fixed product basis, to avoid undesirable effects associated with the fact that the polarized state has very low energy under the system's Hamiltonian.}
\label{tab:expts}
\end{table}

\section{Experimental proposals}

In this section, we present additional details on the two experimental platforms highlighted in the main text, along with four additional platforms that are amenable to implementing our protocol. 
A brief overview of these systems and our proposed implementations is provided in Table~\ref{tab:expts}.

Before discussing the systems individually, we note that a few features are in common in all of the proposals. 
First, we choose the initial state of the protocol to be be quantized along the $X$ direction.
This is motivated by the fact that all of the systems (except the trapped ion quantum computer) feature native interactions that conserve total polarization in the $Z$ basis.
Specifically, we either consider a fully polarized state, or we average over random initial states in the $X$ basis.
The latter approach allows us to circumvent low-energy effects associated with a polarized state; this is most relevant for the two systems with entirely ferromagnetic interactions, i.e.~the dipolar Rydberg atoms and the superconducting qubits.
Moreover, in all cases, we select a ``butterfly'' operator that lies the transverse plane (i.e.~$ V = \sigma_x$ or $S = S_x$).
This generally leads to faster scrambling compared to an operator that overlaps with the conserved quantity $S_z$~\cite{khemani2018operator}.  

\subsection{Dipolar Rydberg atoms}

One particularly suitable platform for realizing our protocol---with either local or global controls---is a quantum simulator based on a 1D or 2D array of atoms 	trapped in optical tweezers \cite{chen2023continuous,bornet2023scalable}.
For each atom, an effective spin-1/2 degree of freedom is encoded in a pair of Rydberg states, which is governed by a long-range XY interaction:
\begin{equation} \label{eq:H-rydberg}
H = -J \sum_{i<j} \frac{a^3}{r^3_{ij}} ( \sigma_x^i \sigma_x^j + \sigma_y^i \sigma_y^j )
\end{equation}
where $J \approx 1$ MHZ is the dipolar interaction strength, $a \sim 10$ $\mu$m is the lattice spacing, and $r_{ij}$ is the distance between atoms.
Crucially, the sign of $J$ is controlled by the specific Rydberg state encoding; for example, $J>0$ occurs for the encoding $\ket{0} = \ket{60S_{1/2},m=1/2}$ and $\ket{1}  = \ket{60 P_{3/2},m=-1/2}$~\cite{gorshkov2011quantum,chen2023continuous}.
Switching between two encodings (via a microwave pulse) allows one to realize time-reversed dynamics. 
Furthermore, one can implement global rotations via microwave pulses and single-site rotations by applying a focused laser beams, which generates a local Stark shift. 
Such control is necessary to prepare a random initial state, as well as to apply the local rotation $e^{i \frac \pi 4  V}$ (for the local control protocol). 
 
In Fig.~\ref{fig:expts_sim}(a), we present simulated results for the protocol with local controls with a 2D array of atoms.
The initial state is randomized over product states in the $X$ basis, and the butterfly operator is $ V = \sigma_x^{N/2}$ (located in the center of the array). 
At early times, we observe a rapid improvement in sensitivity, which for a large system we would expect to follow a quadratic trend, i.e.~$1/\eta \sim t^2$.
At later times, the sensitivity abruptly saturates at $\eta \approx 2/N$, consistent with our prediction from Haar-random evolution and indicating that the system has fully scrambled.

In practice, the improvement in sensitivity over time would compete with the suppression due to accumulation of errors (see Section \ref{sec:errors} for details).
A leading source of decoherence in the system is the lifetime of the Rydberg state, which is typically $T_1 \sim 100$ $\mu$m~\cite{de2018analysis}. 
Based on an interaction strength of $J \sim 1 $ MHz~\cite{chen2023continuous}, we estimate this would enable a high-fidelity preparation of a fully scrambled state for $N \sim 25$ in 1D and $N \sim 100$ in 2D, corresponding to a metrological gain of $8$ and $14$ dB, respectively.

Interestingly, a recent work proposed also proposed the use of 2D Rydberg arrays for generating spin squeezed states~\cite{block2023universal}. 
This approach enables an enhanced sensitivity $N^{-7/10}$ after an evolution time $t \sim N^{2/5}$
With the same 2D array of atoms, we observe that our protocol could obtain a comparable sensitivity at a time $t \sim N^{1/5}$, i.e.~representing a quadratic speedup.
Moreover, the scaling difference between the two protocols is dependent on the dimensionality; whereas the scaling in our protocol improves for higher dimensions, spin-squeezing occurs with the same functional form for all dimensions~\cite{block2023universal}. 
For example, in 1D, the sensitivity for the two protocols would exhibit the same scaling in time, and, in 3D, our protocol would feature a cubic speedup.
This highlights the fact that scrambling occurs at a near-maximal rate under many-body dynamics. 

\subsection{Hybrid spin system in diamond}\label{sec:NV-P1}

\begin{figure}
\includegraphics[width=0.4\linewidth]{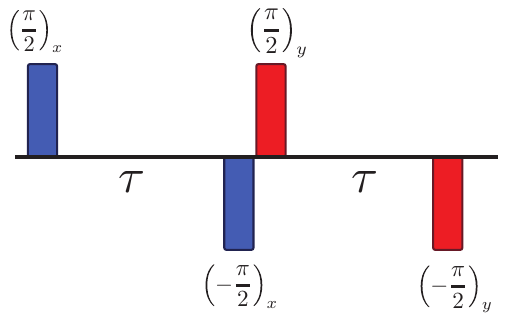}
\caption{A pulse sequence for engineering the hybrid spin Hamiltonian $H$ [Eq.~(\ref{eq:H_NV-P1})] into  $\tilde H^+$ [Eq.~(\ref{eq:H_tilde})]. The sequence consists of two frame rotations with equal duration, $\tau$. In the first, a $\pi/2$ pulse applied along the $X$ direction brings $P_z$ into $P_y$ and $P_y$ into $-P_z$. In the second, a $\pi/2$ pulse applied along the $Y$ direction brings $P_z$ into $P_x$ and $P_x$ into $-P_z$. By rotating in the opposite direction (i.e.~switching $\pi/2$ into $-\pi/2$ and vice versa), the pulse sequence instead generates $\tilde H^-$.}
\label{fig:pulse_engineering}
\end{figure}

As discussed in the main text, our protocol with local controls can naturally be realized in a bulk diamond sample containing two species of electronic spin defects: a relatively high concentration of spin-1/2  nitrogen substitutional defects (P1 centers), and a low density of spin-1 nitrogen-vacancy (NV) centers~\cite{zu2021emergent}.
% 
% Specifically, we consider applying 

% only the $m_s = 0$ and $m_s = -1$ sublevels of the NV center, and we will use pulse engineering to reverse the sign of the interactions~\cite{}.
 
When the two species are off-resonant, the intrinsic magnetic dipole interaction between a single NV center and the surrounding P1 centers gives rise to an effective Hamiltonian~\cite{zu2021emergent}
\begin{align} \label{eq:H_NV-P1}
H &= H_{NV-P1} + H_{P1-P1} \\
H_{NV-P1} &= J_0 \sum_{i} \frac{1}{r_{ij}}(1-3n^z_{ij}) s_z p_z^i \\
H_{P1-P1} &= -\frac{J_0}{2} \sum_{i<j} \frac{1}{r_{ij}}(1-3n^z_{ij}) \left(p_x^i p_x^j + p_y^i p_y^j -2 p_z^i p_z^j \right). \label{eq:H_P1-P1}
\end{align}
Here, $J_0 \approx 52$ MHz-nm$^3$ is the magnetic dipole interaction; $\vec s$, $\vec p^i$ are local spin-1/2 operators acting on the NV center (within the $\ket{m=0} $ and $\ket{m=-1}$ subspace) and individual P1 centers, respectively; $r_{ij}$ is the distance between two defects; and $n_{ij}^z = \hat z \cdot \hat r_{ij}$.

The sign of $H_{NV-P1}$ can easily be reversed by conjugating the evolution via a $\pi$-pulse on the NV center, leading to the effective Hamiltonian
\begin{equation}
H^{-} = -H_{NV-P1} + H_{P1-P1}.
\end{equation}
To reverse the sign of $H_{P1-P1}$, we can apply global pulses to the P1 centers following the pulse sequence shown in Fig.~\ref{fig:pulse_engineering}.
From average Hamiltonian theory, the engineered Hamiltonian in the toggling frame is 
\begin{align} \label{eq:H_tilde}
\tilde H^{\pm} &= \pm \tilde H_{NV-P1} - \frac 1 2 H_{P1-P1} \\
\tilde H_{NV-P1} &= \frac{J_0} {2} \sum_{i} \frac{1}{r_{ij}}(1-3n^z_{ij}) s_z (p_x^i + p_y^i)  ,
\end{align}
where the sign in front of $\tilde H_{NV-P1}$ is determined by direction of the pulses (Fig.~\ref{fig:pulse_engineering}).
Putting these pieces together, we can realize forward and reverse evolution under $\tilde H^+$:
\begin{align}
U &= e^{-i t \tilde H^+} \\
U^\dagger &= e^{-i 2 t \left[  \frac 1 2 \tilde H^{-} + \frac 1 4 (H + H^-)  \right ]} = e^{i t \tilde H^+} .
\end{align}

The full protocol is thus implemented as follows: 
\begin{enumerate}
\item Initialize in a fully polarized state in the $Z$ direction. This is achieved for the NV centers via optical polarization and for the P1 centers via cyrogenic conditions.
\item Rotate the state to the $X$ direction by applying a microwave pulse resonant with ($a$) the two levels of the P1 center, and ($b$) the $\ket{m_s = 0} \leftrightarrow \ket{m_s = -1}$ transition of the NV center.
\item Evolve under the engineered Hamiltonian $\tilde H_+$ by applying the  pulse sequence described above.
\item Apply a local rotation to the NV and evolve backwards under $-\tilde H_+$. The steps up to this point produce the butterfly state.
\item Apply the global sensing signal $e^{-i\phi S_z}$, where $S_z = s_z +  \sum_i p_z^i$. 
\item Evolve forward again under $\tilde H_+$, and measure the polarization of the NV center using optical excitation.
\end{enumerate}

We emphasize that the protocol succeeds despite the presence of strong positional disorder in the spin system: Any position configurations would lead to many-body interactions which produce scrambling behavior and thus an enhancement in sensitivity \footnote{In an ensemble of NV centers, each with their own P1 environment, the total sensitivity $\eta_{\phi=0}$ is given by the average sensitivity over the positional configurations contained in the ensemble. 
% 
% The positional disorder does, however, affect the sensitivity for finite $\phi$. 
% % 
% This is because it contributes additional dephasing to $\langle  V \rangle$ with respect to $\phi$.
}.
Moreover, as discussed in the main text, the scrambling occurs very fast---i.e.~at super-polynomial rate---due to the long-range interactions in three dimensions.
For comparison, a previous scheme has been proposed for entanglement-enhanced sensing with the NV-P1 hybrid system, in which the sensitivity improves linearly in time, i.e.~$1/\eta \sim t$ \cite{goldstein2011environment}. 

\begin{figure}
\includegraphics[width=0.75\linewidth]{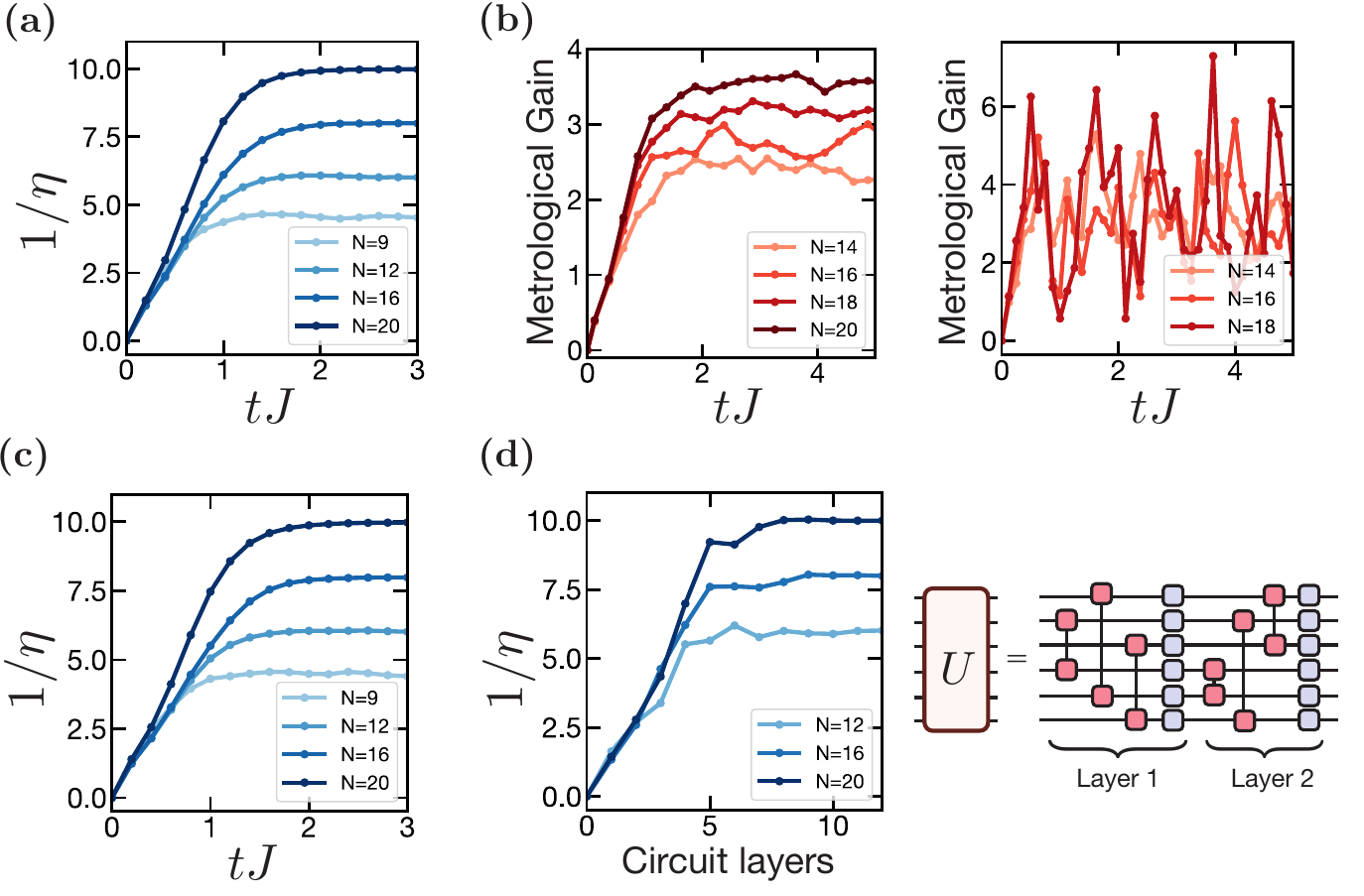}
\caption{Numerical simulations of our protocol with four proposed experimental platforms: (a) Rydberg dipolar atoms in two-dimensions, i.e.~Eq.~(\ref{eq:H-rydberg}) with $r_{ij} = 1$ for nearest neighors; (b) atoms in an optical cativity, i.e.~Eq.~(\ref{eq:H-tree-like}-\ref{eq:J-tree-like}), with (left) $s = -\frac 1 2$ and (right) $s = \frac 1 2$; (c) superconducting qubits with analog interactions, Eq.~(\ref{eq:H-SC-qubits}); (d) trapped ions under a non-local, random unitary circuit. In (a),(c), and (d), the protocol with local controls is performed and the initial state is a random product state in the X basis (averaged over $\sim 10$ realizations). In (b), we implement the protocol with global controls and a fully polarized state initial state, $\ket{\textbf{0}} = \ket{+}^{\otimes N}$. The circuit geometry for the trapped ion simulations is shown in (d). Each layer consists of $N/2$ two-qubit gates (red), acting on random pairs of qubits, and a random single-qubit rotation on each of the qubits (blue).}
\label{fig:expts_sim}
\end{figure}

A particularly attractive feature of electronic spins in diamond is their extremely long intrinsic lifetimes, e.g.~$T_1 \sim$ seconds at low temperatures \cite{jarmola2012temperature}. 
In most settings, the coherence times are instead determined by interactions with other spin defects (both electronic or nuclear spins).
Fortunately, in our protocol, these interactions can be dynamically decoupled by interspersing the pulse sequences described above with global pi-pulses.
As a result, we expect our protocol to be primarily limited by technical constraints, e.g.~the ability to implement pulse sequences with high fidelity and at a pulse rate that is faster than the intrinsic interaction strengths. 
Recent work has led to significant improvements in the robustness of Hamiltonian engineering techniques, which can be leveraged to overcome some of these constraints~\cite{choi2020robust}.

\subsection{Ensemble of NV centers in diamond}\label{sec:NV}

A second system of spin defects which can realize our protocol with \emph{global control} is an ensemble of strongly interacting NV centers \cite{kucsko2018critical}.
The main benefit of this implementation, compared with a hybrid NV-P1 system, is that NV centers can be optically polarized at room temperature, thereby circumventing the need for cryogenic temperatures. 

As in the previous section, time reversal of the NV interactions can be realized by applying pulse engineering techniques.
We first consider the effective interaction between NV centers within the two-level subspace $\{\ket{ 0},\ket{-1}\}$ \cite{kucsko2018critical}:
\begin{equation}
H_0 = J_0 \sum_{i<j} \frac{1}{r_{ij}}(1-3n^z_{ij}) \left(s_x^i s_x^j + s_y^i s_y^j - s_z^i s_z^j \right) ,
\end{equation}
where $\vec s^i$ are spin-1/2 operators acting in the two-level subspace.
Unfortunately, this Hamiltonian \emph{cannot} be time reversed using a sequence of frame rotations.
This results from the observation that the matrix representation of the individual Hamiltonian terms---i.e.~$h_{\mu \nu}$ where $H_{ij} = \sum_{\mu \nu} h_{\mu \nu} s_\mu^i s_\nu^j$ and $\mu,\nu \in \{X,Y,Z\}$---has non-zero trace $\sum_\mu h_{\mu \mu}$, and global operations preserve the trace~\cite{choi2020robust}.

Nevertheless, the NV center contains another potential two-level subspace composed of the $m_s = \pm 1$ sublevels, wherein the effective interaction is
\begin{equation} \label{eq:H_pm1}
H_{\pm 1} = -4 J_0 \sum_{i<j} \frac{1}{r_{ij}}(1-3n^z_{ij}) s_z^i s_z^j.
\end{equation}
Because the trace of $H_{\pm 1}$ has the opposite sign as $H_0$, one can combine the two subspaces to engineer an average Hamiltonian that is traceless.
In particular, alternating between the two subspaces, with duration $\tau$ in the first subspace and duration $\tau/4$ in the second subspace, yields the effective Hamiltonian:
\begin{align} \label{eq:H_avg_NV}
\tilde H &= \frac 4 5 \left (H_0 + \frac 1 4 H_{\pm 1} \right) \\
&= \frac {4 J_0} 5 \sum_{i<j} \frac{1}{r_{ij}}(1-3n^z_{ij}) (s_x^i s_x^j+s_y^i s_y^j-2 s_z^i s_z^j).
\end{align}
This Hamiltonian is simply proportional to the dipolar interaction among P1 centers, i.e.~$H_{P1-P1}$ in Eq.~\ref{eq:H_P1-P1}.
Thus, the same pulse sequence that enabled the time reversal of the P1 interaction, shown in Fig.~\ref{fig:pulse_engineering}, can be applied to the average NV interactions to transform $\tilde H$ into $-\frac 1 2 \tilde H$.

The complete sensing protocol utilizing NV ensembles is outlined as follows:
\begin{enumerate}
\item Optically polarize the NV centers into the $\ket{0}$ state.
\item Rotate the state  via a pi/2 pulse in the subspace $\{\ket{0},\ket{-1}\}$.
\item Evolve under the average Hamiltonian $\tilde H$ by alternating between the $\{\ket{0},\ket{-1}\}$ subspace and $\{\ket{-1},\ket{1}\}$ subspace.
\item Apply a small global rotation to the NV centers, $e^{i \epsilon S_x}$, where the angle $\epsilon$ is optimized as function of the evolution time.
\item Evolve backwards under $-\tilde H$ by (a) alternating between the two subspaces and (b) applying the pulse sequence shown in Fig.~\ref{fig:pulse_engineering} within each subspace. 
\item Apply the global sensing signal $e^{-i\phi S_z}$, where $S_z = \sum_i s_z^i$. 
\item Evolve forward again under $\tilde H$, and measure the total polarization of the NV centers via optical excitation.
\end{enumerate}

To our knowledge, implementing time reversal by applying pulse engineering sequences to \emph{multiple subspaces} has not been previously proposed. 
Successfully demonstrating this technique, which we refer to as ``subspace engineering'', and comparing it to more standard pulse sequences involving a single subspace, represents an exciting experimental prospect.

% The main challenge in this approach is to reverse the dipolar interaction.

\subsection{Atoms in an optical cavity}

One of the most successful platforms for demonstrating entanglement-enhanced sensing consists of atoms coupled in an optical cavity \cite{pedrozo2020entanglement,colombo2022time,li2023improving}.
Conventionally, this enhancement is achieved by evolving under collective large-spin interactions (e.g.~the one-axis twisting Hamiltonian) to generate a squeezed state. 
However, this approach does not succeed for more general types of dynamics, arising in e.g.~a multi-mode cavity \cite{vaidya2018tunable} or via programmable interactions \cite{periwal2021programmable}.  
 
We consider the latter setting with a system of spin-1/2 atoms.
The programmable spin-exchange interactions are described by an effective Hamiltonian \cite{periwal2021programmable}:
\begin{equation} \label{eq:H-tree-like}
H = \sum_{i<j} J(r_{ij}) \left( \sigma_x^i \sigma_x^j + \sigma_y^i \sigma_y^j \right),
\end{equation}
where the sign and magnitude $J(r_{ij})$ are controllable by laser drives.
Motivated by Ref.~\cite{bentsen2019treelike}, we select the interaction strength to be of the form 
\begin{equation} \label{eq:J-tree-like}
J(r_{ij}) = \left\{
    \begin{array}{ll}
        (-1)^n |i-j|^s & \; \text{if } |i-j|=2^n, n \in \mathbb{Z}\\
        0 & \; \textrm{otherwise}
    \end{array}
\right .
% 
% \left( \sigma_x^i \sigma_x^j + \sigma_y^i \sigma_y^j \right)
\end{equation}
where the parameter $s$ interpolates between a quasi-one-dimensional geometry ($s < 0$) and a tree-like geometry ($s > 0$).
Note that we include a mix of ferromagnetic and anti-ferromagnetic couplings. 
This is anticipation of initializing to protocol with a fully polarized initial state, $\ket{\textbf{0}} = \ket{+}^{\otimes N}$; the anti-ferromagnetic interactions serve to raise the temperature this state.

Although local control is theoretically possible in this setup~\cite{swingle2016measuring}, it is most natural to realize our protocol with global controls. 
Numerical simulations for the sensitivity as a function of evolution time are depicted in Fig.~\ref{fig:expts_sim}. 
For $s < 0$, we observe that the sensitivity quickly reaches a saturation value that improves with system size $\sim N$, indicating a Heisenberg-like enhancement.
% 
% Intriguingly, the linear prefactor is slightly worse than our expectations from the Haar-random calculation, indicating that the system does not reach a fully scrambled state. 
% 
Intriguingly, in the case of $s > 0$, we find a qualitatively different behavior: the sensitivity exhibits large fluctuations and does not improve systematically with system size. 
This suggests that the tree-like geometry does not lead to fully scrambling behavior.
Understanding the subtle interplay between many-body dynamics and improved sensitivity for such non-trivial geometries would be an interesting future direction.

\subsection{Superconducting qubits with analog interactions}\label{sec:SC}

Tremendous progress has been made in developing quantum processors based on 2D arrays of superconducting transmon qubits \cite{arute2019quantum,kim2023evidence,braumuller2022probing}.
While such processors are often controlled with digital gates, we consider an implementation of our protocol which utilizes the intrinsic (analog) interactions between tunable-frequency qubits.
In particular, when the qubits are brought on resonance, their interaction is described by a local XY model~\cite{braumuller2022probing,andersen2024thermalization},
\begin{equation} \label{eq:H-SC-qubits}
H = J \sum_{\langle i,j \rangle} \left( \sigma_x^i \sigma_x^j + \sigma_y^i \sigma_y^j \right),
\end{equation}
where the coupling strength $|J|$ is typically $10-100$ MHz \cite{foxen2020demonstrating}.
As demonstrated in Ref.~\cite{braumuller2022probing}, when the lattice of qubits is bipartite, the sign of interaction can be quite easily reversed by conjugating the evolution with $\pi$-pulses, i.e.~$-H = \left(\prod_{i\in \mathcal{S}} \sigma_x^i\right) H \left(\prod_{i\in \mathcal{S}} \sigma_x^i \right)$, where $\mathcal{S}$ is one part of the bipartite lattice.
This reversibility, in addition to local rotations generated by microwave pulses, enables the realization of our protocol with local control.

In Fig.~\ref{fig:expts_sim}, we show numerical results for our protocol in a 2D array of up to $20$ qubits.
While the functional form of the early-time growth cannot be discerned, at larger sizes one expects the nearest-neighbor interactions to lead to ballistic growth of the form $\sim (Jt)^2$.
Based on an estimate of this growth rate, a qubit lifetime of $T_1 \sim 20 \mu$s \cite{rosenberg2023dynamics}, and a coupling strength $J\sim 50$ MHz \cite{foxen2020demonstrating}, we predict that a fully scrambled state can be prepared with $\sim 400$ qubits, leading to a metrological gain of 20 dB.
This would significantly surpass the current record for metrological gain of 11.8 dB, recently demonstrated via atoms in an optical cavity~\cite{colombo2022time}.

For comparison, a more conventional approach for obtaining a metrological gain in a digital quantum processor would be to prepare a GHZ state (the current record is a GHZ state with 60 qubits and a fidelity of $0.59$) \cite{bao2024schr,mooney2021generation}.
It is natural to consider which of these strategies would lead to a larger metrological gain on realistic devices. 
On the one hand, the theoretical sensitivity for a GHZ state at equivalent sizes is a factor of two better than our protocol. 
Additionally, sensing based on a GHZ state requires two layers of entangling gates (i.e.~the state preparation circuit and its inverse, assuming that robustness to noise is desired), whereas our protocol requires three steps of many-body evolution.
% % 
On the other hand, our protocol is much more robust against control errors, since it does not require precisely calibrated two-qubit gates; indeed, such errors often represent a significant fraction of the total error~\cite{mi2021information}. 
Moreover, the total evolution time for implementing our protocol may be shorter, since the analog interactions are ``always on'', thereby reducing the effect of decoherence~\cite{andersen2024thermalization}.
Testing these advantages in practice would be of tremendous interest and may provide a useful tool for benchmarking large-scale quantum processors.

\subsection{Trapped-ion quantum computer}\label{sec:traped-ion}

Lastly, we consider an implementation of protocol on a trapped-ion quantum computer~\cite{pogorelov2021compact,egan2021fault}.
Unlike the previous proposals which rely on analog evolution, we utilize discrete quantum gates to generate the many-body unitary $U$.
Specifically, we construct circuits with interspersed layers of two-qubit and single-qubit gates.
For the two-qubit gates, we choose $N/2$ pairs of qubits at random and apply the  native Molmer-Sorensen interaction, $e^{i \frac \pi 4 \sigma_i^x \sigma_j^x}$, to each pair. 
This arrangement takes advantage of the all-to-all connectivity of trapped ions. 
For the single-qubit gates, we apply $e^{i\alpha_i \sigma_z^i} e^{i\frac \pi 4 \sigma_y^i} e^{i\beta_i \sigma_z^i}$, where $\alpha_i, \beta_i \in [0,2\pi]$ are chosen from a uniform distribution. 
In Fig.~\ref{fig:expts_sim}(d), we plot numerical results for the sensitivity as function of circuit depth using the sensing protocol with local controls. 
Much like our previous results with analog evolution, we observe an initial rise in sensitivity, followed by saturation at $\eta \approx 2/N$.
Owing to the all-to-all connectivity, the circuit depth to reach saturation scales favorably with system size; indeed, one expects it to occur in $\sim \log N$ layers at large system sizes. 

% , i.e.~a linear rise in sensitivity, followed by saturation at $\eta = 2/N$\footnote{We note that the 1D geometry is shown as a proof of principle. In practice, one could easily modify the circuit to saturate in a smaller number of layers by implementing long-range entangling gates, thereby taking advantage of the all-to-all connectivity of the trapped-ion platform.}. 

As discussed in the main text, an important feature of our protocol is its robustness against coherent errors, which are considered to be a dominant error source in trapped-ion systems.
Physically, such errors arise from low-frequency fluctuations in the laser drive amplitudes, causing imperfections in the rotation angles, i.e.~$\theta_{ij} \rightarrow \theta_{ij}^\prime$.
These errors will limit the ability to prepare finely-tuned metrological states, including a GHZ state.
However, in our protocol, if these errors can be time reversed they have essentially no impact on the achievable sensitivity---they would simply adjust the many-body unitary, $U \rightarrow U^\prime$, and, at late times, this would still result in a fully scrambled state. 
Thus, at large system sizes and / or high coherent error rates, we expect our protocol to provide a larger metrological gain compared to sensing based on a GHZ state.

\section{Details on stochastic model used for large-scale numerical studies}

In this section, we introduce a stochastic model for operator growth dynamics which allows us to predict the large-scale behavior of our protocol for the two proposed systems of spin defects.
The model is inspired by previous work on quantum information scrambling, where it has been argued that growth of operators under \emph{long-range} Hamiltonian dynamics can be qualitatively captured at long timescales by stochastic transitions \cite{chen2019quantum,zhou2020operator,zhou2023operator}.
For our purposes, we model these transitions using Haar-random gates and determine the probability of each gate based on the strength of the spin interactions.

In more detail, consider a Hamiltonian composed of two-body interactions, $H = \sum_{ij\mu} h^\mu_{ij}$, where $h^\mu_{ij}$ indicates a particular two-body operator acting on qubits $i$ and $j$. 
Our model consists of mapping the analog evolution $U=e^{-iHt}$ to a circuit composed of $D$ time steps, $\tilde U = U_D \cdots U_2 U_1 $.
In each time step, we apply a set of two-qubit, Haar-random gates, where the probability of a gate occuring between qubits $i$ and $j$ is $P_{ij} = \delta t \sum_\mu |h^\mu_{ij}|^2$. 
We set $\delta t \ll 1/J_{\textrm{typ}}$, where $J_{\textrm{typ}}$ is the typical interaction strength, such that there is a low density of gates per time step.

A key feature of Haar-random gates is that, for measuring certain quantities, the average over Haar-random gates is equivalent to the average over random Clifford gates \cite{nahum2018operator}.
The average sensitivity of our protocol (with either global or local control) represents such a quantity; this results from the fact that the sensitivity contains three copies of $U$ and $U^\dagger$, and Clifford unitaries form a 3-design for qubits \cite{webb2015clifford,kueng2015qubit,zhu2017multiqubit}.
Thus, the average sensitivity of the stochastic model can be computed efficiently using Clifford numerics.

We apply the stochastic model to predict the sensitivity of our protocol for either local or global control. 
For the local protocol, we compute the sensitivity $\eta_{\phi=0}$ by measuring the average polarization density $P(S_z)$ of $\tilde U^\dagger  V \tilde U \ket{\textbf{0}}$ [see Eq.~(\ref{eq:avg_size})].
This is easily accomplished by evolving $ V$ in the Heisenberg picture with Clifford gates, and then counting the number of $\sigma_x$ and $\sigma_y$ operators within the Pauli string $ V(t)$.

Computing the sensitivity for the global protocol is somewhat more involved.
We begin by expressing the mean outcome as
\begin{equation}
\langle S_x \rangle_{\phi} = \bra{\textbf{0}} U^\dagger e^{i\epsilon S_x} U e^{i\phi S_z} U^\dagger S_x U e^{-i\phi S_z} U^\dagger e^{-i\epsilon S_x} U \ket{\textbf{0}},
 \end{equation} 
where $\vec S = \frac{1}{2} \sum_i \vec \sigma^i$ are global spin-1/2 operators.
To proceed, we expand the operator $S_x$ in a Pauli basis:
\begin{equation}
\langle S_x \rangle_{\phi}  = \frac{1}{2} \sum_{a,b \in \left\{0,1\right\}^N}\sum_{i=1}^N c_{a} c_{b}^* \bra{\textbf{0}} U^\dagger P_{a} U e^{i\phi S_z} U^\dagger \sigma_x^i U e^{-i\phi S_z} U^\dagger P_{b} U \ket{\textbf{0}},
 \end{equation} 
where $P_{a} = \prod_{i \in a} \sigma_x^i$ contains $\sigma_x^i$ on all sites for which $a$ is non-zero, $c_a = (i \sin (\epsilon/2))^{|a|}(\cos(\epsilon/2))^{N-|a|}$, and $|a|= \sum_i a_i$ is the Hamming weight of $a$.
For a Clifford unitary $U$, the state $U^\dagger P_b U \ket{\textbf{0}}$ is an eigenstate of $S_z$.
This allows us to pull the factors of $e^{\pm i \phi S_z}$ outside the expectation value, giving
\begin{equation}
\langle S_x \rangle_{\phi} = \frac{1}{2} \sum_{a,b \in \left\{0,1\right\}^N}\sum_{i=1}^Nc_{a} c_{b}^* e^{i\phi_{ab}} \bra{\textbf{0}} U^\dagger P_{a}  \sigma_x^i  P_{b} U \ket{\textbf{0}},
 \end{equation} 
where $\phi_{ab} = \phi (\mathcal{S}_z \{U^\dagger P_a U\} - \mathcal{S}_z \{U^\dagger P_b U\})$ and $\mathcal{S}_z\{P\} = $ \# of $\sigma_x$,$\sigma_y$ in $P$.
With high probability, the matrix element $\bra{\textbf{0}} U^\dagger P_{a} \sigma_x^i  P_{b} U \ket{\textbf{0}}$ is non-zero if and only if $P_{a}  \sigma_x^i  P_{b} = \mathds{1}$.
With this simplification, we have
\begin{equation}
\langle S_x \rangle_{\phi} \approx \frac{1}{2}  \sum_{a \in \left\{0,1\right\}^N}\sum_{i=1}^N \left|c_{a}\right|^2 (i\tan(\epsilon/2))^{|b_i|-|a|} e^{i\phi_{ab_i}} ,
 \end{equation} 
where $b_i$ differs from $a$ only on the $i^{\text{th}}$ bit.
We note that $|c_a|^2$ is the probability of sampling $|a|$ from a binomial process with $N$ draws of probability $\sin^2(\epsilon/2)$, and thus we can approximate the above expression using Monte Carlo sampling.

In summary, the full procedure for estimating $\langle S_x \rangle_\phi$ is as follows:
\begin{enumerate}
\item Sample $a$ from a binomial distribution and $i$ from a uniform distribution. The bitstring $b_i$ is immediately given by flipping $a$ on the $i^{\text{th}}$ bit.
\item Compute $\phi_{ab} = \phi (\mathcal{S}_z \{U^\dagger P_a U\} - \mathcal{S}_z \{U^\dagger P_b U\})$ by time evolving $X_a$ and $X_b$ under a Clifford circuit.
\item Average the quantity $(i\tan \epsilon)^{|b_i|-|a|} e^{i\phi_{ab_i}}$ over many samples.
\end{enumerate}
We estimate the sensitivity via $\eta^{-1}_{\phi=0} \equiv (\partial_\phi \langle S_x \rangle_\phi/\Delta S_{x,\phi})_{\phi = 0} \approx \langle S_x \rangle_\phi/\sqrt{N}$, where $\phi \ll 1$, and for circuits with Haar-random gates, on average, $\langle S_x \rangle_{\phi=0}$ and $(\Delta S_x)_{\phi=0} = \sqrt{N}/2$.

\begin{figure}[t]
\includegraphics[width=0.8\linewidth]{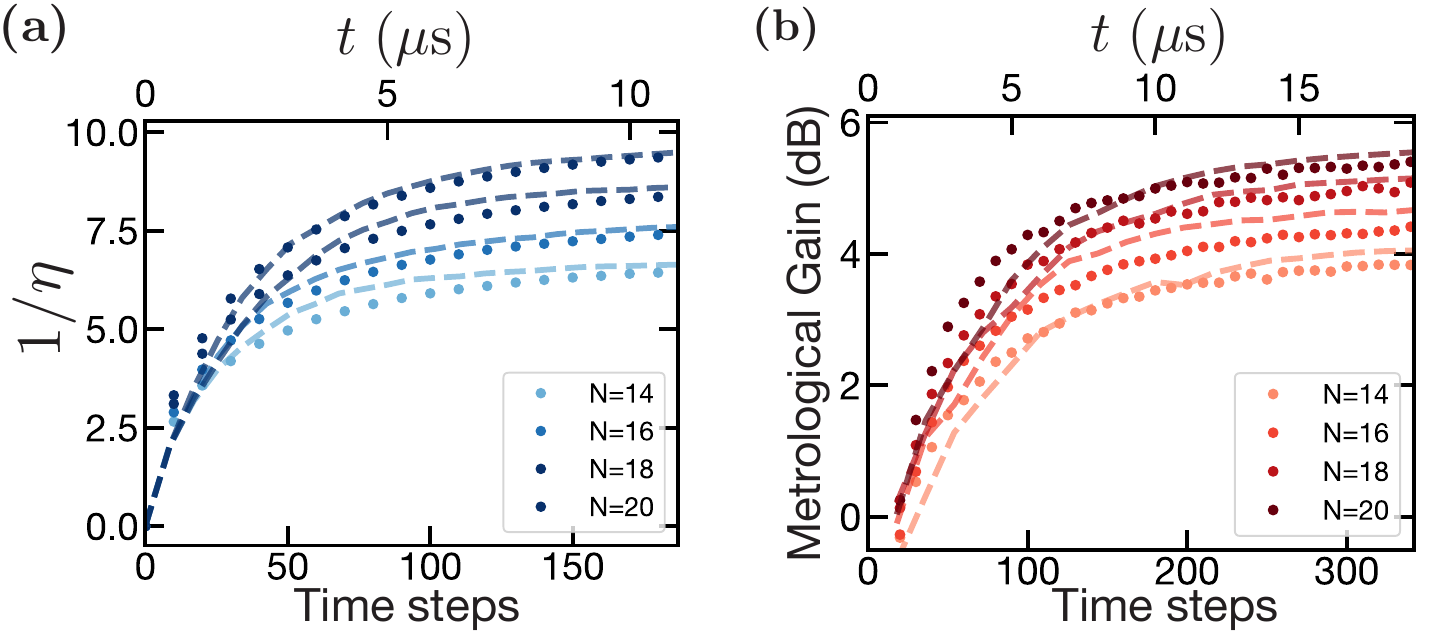}
\caption{Comparison between the stochastic model (solid lines) and exact dynamics (dashed) for the two proposed spin models at small sizes. (a) The sensitivity of the local protocol applied to the hybrid spin system with single NV center surrounded by a cluster of P1 centers. The density of spin defects is 100 ppm dynamics, and their interactions are governed by $\tilde H^+$ in Eq.~\ref{eq:H_tilde}. (b) The metrological gain of the global protocol applied to an ensemble of NV centers. The density of the NV centers is 100 ppm, and their dynamics are governed by $\tilde H$ in Eq.~\ref{eq:H_avg_NV}. The results for the stochastic model are plotted as a function of the number of discrete time steps, and the results for the exact dynamics are plotted with respect to continuous time evolution. By comparing the results, we estimate the conversion factor between discrete steps and continuous evolution time. The results for the stochastic model are averaged over $\sim 10^4$ realizations  (including different positional configurations and Clifford circuits), and the results for the exact dynamics are averaged over $\sim 10$ positional configurations.}
\label{fig:RUC_benchmarking}
\end{figure}

To benchmark the stochastic model, we calculate the sensitivities for up to $N = 20$ spins for  ($i$) the local protocol based the interactions of the hybrid spin system, and ($ii$) the global protocol based on the interactions of an ensemble of NV centers.
In ~Fig.~\ref{fig:RUC_benchmarking}, we compare these results to the the sensitivities obtained from simulations with exact dynamics. 
The two methods yield good qualitative agreement for both the local and global protocols.
Moreover, by matching the growth rate at early times, we can estimate the conversion factor between discrete time steps in the stochastic model and evolution time for the exact dynamics.
Utilizing this conversion factor, we simulate the behavior of much larger spin systems, as shown in the insets of Fig.~4 of the main text.
We expect the results of the stochastic model to provide a good approximation for the exact behavior of the experimental system, in a regime that would be intractable to simulate with the exact dynamics.

\section{Detailed analysis of the effects of noise and decoherence}\label{sec:errors}

We now provide a detailed accounting of the effect of experimental errors on our sensing protocols.
We begin with a brief discussion of readout and initialization errors, which as mentioned in the main text, decrease the sensitivity by only a constant factor.
We then turn to incoherent errors during time evolution, and, borrowing from the results of Ref.~\cite{schuster2022operator}, derive the suppression factor indicated in the main text.

Readout errors have a particularly small effect on our protocol.
For the local control protocol, a local readout error rate $\gamma_r$ suppresses the expectation value of $V$ by a constant factor, $\langle V \rangle_\phi \rightarrow (1-\gamma_r) \langle V \rangle_\phi$.
A similar suppression occurs for the global protocol, $\langle S_x \rangle_\phi \rightarrow (1-\gamma_r) \langle S_x \rangle_\phi$, since $S_x$ is a sum of single-body operators.
In both cases, the sensitivity is suppressed by the same factor, $\eta^{-1}_\phi \rightarrow (1-\gamma_r) \eta^{-1}_\phi$.

To address initialization errors, consider performing the protocol with an initial density matrix $\rho$ instead of $\dyad{\textbf{0}}$.
We denote the mean polarization of $\rho$ as $\tr( \rho S_z ) = (1-\gamma_i) N/2$, where $\gamma_i$ quantifies the local initialization error rate.
We also suppose that the polarization distribution of $\rho$ has width $\lesssim \sqrt{N}$, which is appropriate for local initialization errors.
After butterfly state preparation, the density matrix becomes
\begin{equation}
    \left( \frac{\mathbbm{1}+i V(t)}{\sqrt{2}} \right) \rho \left( \frac{\mathbbm{1}-i V(t)}{\sqrt{2}} \right) = \frac{1}{2} \left( \rho +  V(t) \rho V(t) + i \left[ V(t) \rho - \rho V(t) \right] \right),
\end{equation}
where $V(t) = U^\dagger V U$.
The first and second terms correspond to the two trajectories of the butterfly state, and the third term to the coherence between them.
As in the error-free case, for small angles $\phi \lesssim 1/\sqrt{N}$ the rotation $e^{- \phi S_z}$ simply applies an overall phase to each trajectory of the butterfly state.
Working in the late time regime where $e^{i \phi S_z} V(t) \rho \approx V(t) \rho$, this leads to the density matrix
\begin{equation}
    e^{i \phi S_z} \left( \frac{\mathbbm{1}+i V(t)}{\sqrt{2}} \right) \rho \left( \frac{\mathbbm{1}-i V(t)}{\sqrt{2}} \right) e^{-i \phi S_z}  \approx \frac{1}{2} \left( \rho +  V(t) \rho V(t) + i \left[ e^{-i \phi (1-\gamma_i) N/2} V(t) \rho - e^{i \phi (1-\gamma_i) N/2} \rho V(t) \right] \right).
\end{equation}
As in the error-free case, only the third term (in square brackets) will contribute to the final expectation value of $V$.
Applying the final unitary and taking the expectation value gives
\begin{equation}
    \langle V \rangle_\phi  \approx \sin \left( \phi \, \frac{(1-\gamma_i) N}{2} \right).
\end{equation}
Taking the derivative with respect to $\phi$, we see that the sensitivity is decreased by only a constant factor relative to the error-free case, $\eta^{-1}_\phi \rightarrow (1-\gamma_i) \eta^{-1}_\phi$.

\begin{figure}[t]
\includegraphics[width=0.8\linewidth]{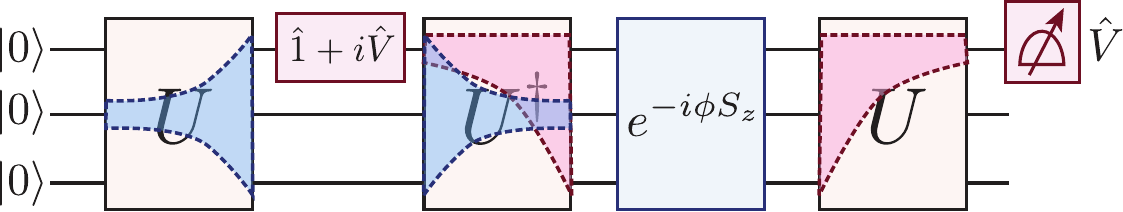}
\caption{Illustration of the effect of errors on our sensing protocol. Errors within the light-cone of ${V}$ (red dashed areas) suppress the coherence between the two trajectories of the butterfly state. Errors within the light-cone of a local polarization operator $\sigma^z_i$ (blue dashed areas, shown for a representative polarization operator) suppress the polarization of the first trajectory of the butterfly state. The sensitivity of the protocol is affected by both types of errors, and is thus suppressed proportional to the local error rate $\gamma$ multiplied by the space-time volume of the four light-cones. Note that initialization and readout errors can be included in this diagram as well; since they only act at times when the light-cones have size one, they have only an $\mathcal{O}(1)$ effect on the sensitivity.}
\label{fig:error}
\end{figure}

We now turn to incoherent errors during time evolution.
As mentioned in the main text, incoherent errors have two effects on the protocol: they suppress the mean polarization in the first trajectory of the butterfly state, and they suppress the coherence between the two butterfly trajectories.
Both effects suppress the sensitivity, the first by suppressing the first derivative of $\langle {V} \rangle_\phi$ with respect to $\phi$ (similar to initialization errors), and the second by suppressing the overall magnitude of $\langle {V} \rangle_\phi$ (similar to readout errors).
However, unlike initialization and readout errors, errors during time evolution occur when the state is highly-entangled and thus have a stronger effect (Fig.~\ref{fig:error}).% 

To explore this in more detail, let us replace the unitary evolution $\rho \rightarrow U \rho U^\dagger$ by evolution under a noisy quantum channel, $\rho \rightarrow \mathcal{E}_\gamma \{ \rho \}$.
For analog evolution, the quantum channel might be generated by a Lindbladian, $\mathcal{E}_\gamma = e^{\mathcal{L}_\gamma t}$, where $\mathcal{L}_\gamma$ includes both Hamiltonian evolution and local noise operators with strength $\gamma$.
For digital evolution, the quantum channel might correspond to a sequence of unitary gates interspersed with local noise channels of strength $\gamma$.
In any case, we will assume that the quantum channel corresponding to $U^\dagger$ is the \emph{conjugate} of the channel corresponding to $U$, defined via $\tr \big( M \cdot \mathcal{E}_\gamma \{ \rho \} \big) = \tr \big( \mathcal{E}_\gamma^\dagger \{ M \} \cdot \rho \big)$\footnote{On a technical level, this requires assuming that the noise is unital. We expect non-unitality of the noise channel to contribute at sub-leading order in $\gamma$; see the supplemental material of Ref.~\cite{schuster2022operator} for a full discussion.}.
This reduces to standard time reversal when the evolution is unitary.

Let us now analyze the sensitivity of the protocol at $\phi = 0$.
We have
\begin{equation}
    \partial_\phi \langle V \rangle_\phi = \text{Im} \left[ \tr( S_z\cdot  \mathcal{E}_\gamma^\dagger \{ V \} \cdot \mathcal{E}_\gamma^\dagger \left\{ \frac{\mathbbm{1}+iV}{\sqrt{2}} \mathcal{E}_\gamma \left\{ \dyad{\textbf{0}} \right\} \frac{\mathbbm{1}-i V}{\sqrt{2}} \right\}  ) \right],
\end{equation}
where we apply (the conjugate of) the final time evolution to the measurement operator $V$ instead of the quantum state.
As in the error-free case, we can drop terms that contain an odd number of $ V$.
Moreover, terms where a single $V$ operator appears in between $S_z$ and the initial state $\dyad{\textbf{0}}$ can also be dropped if the system is fully scrambled after application of $\mathcal{E}_\gamma$.
These correspond to the polarization of the second trajectory of the butterfly state or, in other language, to OTOCs that have decayed to zero.
Dropping these terms gives
\begin{equation}
    \partial_\phi \langle V \rangle_\phi \approx \tr( S_z\cdot  \mathcal{E}_\gamma^\dagger \{ V \} \cdot \mathcal{E}_\gamma^\dagger \left\{ V \cdot \mathcal{E}_\gamma \left\{ \dyad{\textbf{0}} \right\} \right\}  ).
\end{equation}
Now, note that the initial state $\dyad{\textbf{0}}$ can be decomposed as a sum of stabilizers as
\begin{equation}
    \dyad{\textbf{0}} = \frac{1}{2^N} \sum_{s \in \{0,1\}^N}\bigotimes_{i=1}^N (\sigma^z_i)^{s_i}.
\end{equation}
To good approximation, only the single-body stabilizers $\sigma^z_i$ contribute to the expectation value, since they can ``contract'' with the same stabilizer in $S_z$.
Keeping only these stabilizers, we have
\begin{equation} \label{eq: final approximation error}
    \partial_\phi \langle V \rangle_\phi \approx \frac{1}{2} \sum_{i=1}^N \frac{1}{2^{N}} \tr( \sigma^z_i \cdot  \mathcal{E}_\gamma^\dagger \{ V \} \cdot \mathcal{E}_\gamma^\dagger \left\{ V \cdot \mathcal{E}_\gamma \left\{ \sigma^z_i \right\} \right\}  ),
\end{equation}
which is our final approximation.
The approximation resembles a ``doubled'' version of the Loschmidt echo, which depends on both the fidelities of a local operator $\sigma^z_i$ at time zero and a local operator $V$ at time $t$.

To understand how this approximation depends on the local noise rate $\gamma$, we invoke the results of Ref.~\cite{schuster2022operator}.
There, it was argued that for ergodic many-body quantum dynamics, the decay of the Loschmidt echo,
\begin{equation}
    \mathcal{N}_\gamma( M ) = \frac{1}{2^N} \tr( M \cdot \mathcal{E}_\gamma^\dagger \{ \, \mathcal{E}_\gamma \{ M \} \, \} ),
\end{equation}
is controlled by the effective space-time volume of the time-evolved operator $M$,
\begin{equation} \label{eq: Loschmidt approx}
    \mathcal{N}_\gamma( M ) \approx \exp \left( -2\gamma \, \text{Vol}\left[M(0\rightarrow t) \right] \right).
\end{equation}
Here, the space-time volume is defined as the integral over time of the \emph{size} of the operator ${M}$, 
\begin{equation}
    \text{Vol} \left[M(0\rightarrow t) \right] = \int_0^t dt' \mathcal{S}(t'),
\end{equation}
where the size is given by the (average) number of qubits that $M$ acts upon,
\begin{equation}
    \mathcal{S}(t') = \sum_P | c_P(t') |^2 \mathcal{S}_P,
\end{equation}
where $ M(t') = \sum_P c_P(t')$ is the Pauli decomposition of $M$ at time $t'$ and $\mathcal{S}_P = (\# \text{ of } \sigma_x, \sigma_y, \sigma_z \text{ in } P)$ is the weight of the Pauli operator $P$.
In principle, one should compute this volume for time evolution under the noisy quantum channel~\cite{schuster2022operator}. However, to estimate the leading order dependence in $\gamma$ we can substitute the volume under unitary evolution.
In short-range interacting systems, the space-time volume is proportional to the volume of the operators' light-cone, and Eq.~(\ref{eq: Loschmidt approx}) simply states that only errors within the light-cone contribute to the decay of the Loschmidt echo.

We can straightforwardly apply this approximation to the sensitivity in Eq.~(\ref{eq: final approximation error}).
The first quantum channel (applied to $\sigma^z_i$) contributes a factor $\gamma \text{Vol}[ \sigma^z_i(0 \rightarrow t) ]$ to the exponent.
The final quantum channel (applied to $V$) contributes a factor $\gamma \text{Vol}[ V(0 \rightarrow t) ]$.
The only subtlety is the middle quantum channel (applied to $V \cdot \mathcal{E}_\gamma \{ \sigma^z_i \}$).
This contributes a factor proportional to the space-time volume of the \emph{product} of $\sigma^z_i$ and $V$, where the former is local at the end of the evolution and the latter at the beginning.
In a slight abuse of notation, we denote this quantity as $\text{Vol}[ \sigma^z_i(t \rightarrow 0) \cup V(0 \rightarrow t)]$.
Putting it all together, we estimate
\begin{equation} \label{eq: final approximation error}
    \partial_\phi \langle V \rangle_\phi \approx \frac{1}{2} \sum_{i=1}^N \exp(-\gamma \left( \text{Vol} \big[ \sigma^z_i(0 \rightarrow t) \big] + \text{Vol}\big[  V(0 \rightarrow t) \big] + \text{Vol}\big[ \sigma^z_i(t \rightarrow 0) \cup V(0 \rightarrow t)\big] \right) ).
\end{equation}
The relevant light-cones are depicted visually in Fig.~\ref{fig:error}.
Note that the volume of the product will (typically) be upper bounded by the sum of the individual volumes.
This gives a lower bound on the sensitivity,
\begin{equation}
\begin{split}
    \partial_\phi \langle V \rangle_\phi & \gtrsim \left( \frac{1}{2} \sum_{i=1}^N \exp \left(-2\gamma \text{Vol} \big[ \sigma^z_i(0 \rightarrow t) \big] \right) \right) \cdot \exp(-2\gamma \text{Vol}\big[ V(0 \rightarrow t) \big] ) \\
    & = \frac{N}{2} \cdot \overline{\mathcal{N}_\gamma (\sigma^z_i)} \cdot \mathcal{N}_\gamma(V)
\end{split}
\end{equation}
The first term is given by $N/2$ multiplied by the average Loschmidt echo  of $\sigma^z_i$. This corresponds to the loss in polarization of the first butterfly trajectory.
The second term is given by the Loschmidt echo of $V$, and corresponds to the loss of coherence between the first and second trajectories of the butterfly state.

Let us consider this final expression for a system with local interactions in $d$ dimensions. 
We assume local operators spread ballistically with a butterfly velocity $v_B$, resulting in an operator size $S(t) \approx (v_B t)^d$ and a volume $\textrm{Vol}[V(0\rightarrow t)] \approx \textrm{Vol}[\sigma_i^z(0\rightarrow t)] \approx \frac 1 {(d+1)v_B} (v_B t)^{d+1}$.
To reach a fully scrambled state, we take $t \approx N^{1/d}/v_B$.
This leads to an overall sensitivity,
\begin{equation}
\partial_\phi \langle V \rangle_\phi \gtrsim \frac N 2 \textrm{exp}\left(-\frac{4}{d+1} \frac \gamma {v_B} N^{\frac{d+1}{d}}\right).
\end{equation}
We utilize this expression to estimate the sensitivity under decoherence for the dipolar Rydberg atoms and superconducting qubits.

\begin{figure}
\includegraphics[width=0.82\linewidth]{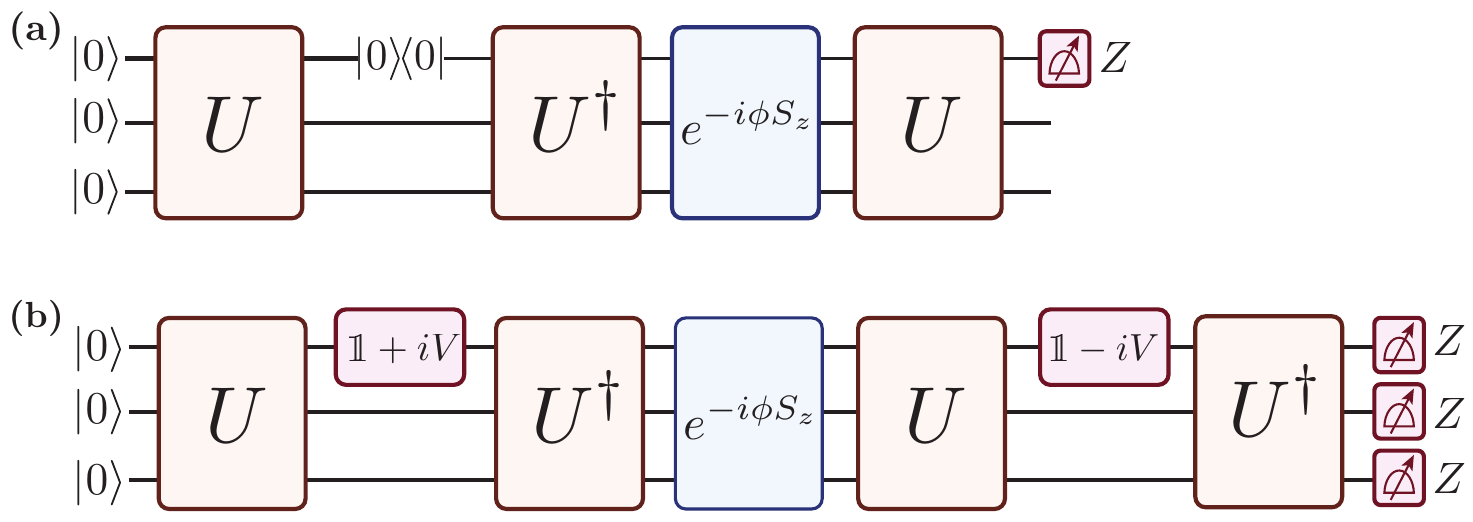}
\caption{Two protocols for measuring the \emph{real} part of $\Phi(\phi)$. (a) The first protocol is identical to the original protocol with local controls [Fig.~1(a) of the main text], except we replace the local rotation, $e^{i\frac pi 4 V} \sim \mathbbm{1} + i  V$, with a local projection,~$\dyad 0 \sim \mathbbm{1} + Z$. (b) The second protocol, which we dub a ``double echo'', involves applying the butterfly state preparation circuit and its inverse. The final state is measured in the computational basis, and either the return probability to the initial state or the average polarization is computed. This approach may be applied with either local or global controls to prepare the butterfly state.}
\label{fig:opposite_quad}
\end{figure}

\section{Experimental protocol to measure the opposite quadrature} \label{sec: opposite quadrature}

As discussed in the previous section, the protocols shown in the main text (i.e.~Fig.~1 and 3) are directly related to the imaginary part of the characteristic function $\Phi(\phi)$ of the polarization distribution $P(S_z)$.
This implies that they achieve a high sensitivity only at certain values of $\phi$, including, most notably, $\phi = 0$.
In order to maintain a high sensitivity over a continous range of $\phi$, it is necessary to measure the opposite quadrature---i.e.~the real part of the characteristic function.

There are two straightforward modifications of our protocol that achieve this goal, depicted in Fig.~\ref{fig:opposite_quad}.
The first approach, which applies only to the protocol with local controls, is to replace the local rotation by a \emph{projection}, e.g.~$(1+{V})/2$ for a Pauli operator $V$.
The measurement outcome then becomes
\begin{equation} \label{eq:V_real}
\begin{split}
\langle V \rangle_\phi = &\frac 1 2 \bra{\textbf{0}} V(t) \ket{\textbf{0}} - \frac 1 2 \bra{\textbf{0}} V(t) e^{i \phi S_z} V(t) e^{-i \phi S_z} V(t) \ket{\textbf{0}} \\ 
& + \textrm{Re}\left [ e^{i \phi \frac N 2} \bra{\textbf{0}} V(t) e^{-i \phi S_z} V(t) \ket{\textbf{0}} \right ] \\
= & \textrm{Re}\left [ e^{i \phi \frac N 2} \Phi(\phi) \right ],
\end{split}
\end{equation}
where $\Phi(\phi)$ is defined below Eq.~(\ref{eq:im_Phi}), and, in the second line, we assume that the first two terms have vanishing expectation values\footnote{The first two terms may also be directly cancelled by projecting onto the opposite state, i.e.~using $(1-V)/2$, and measuring $-V$. Averaging this outcome with Eq.~(\ref{eq:V_real}) leaves only the final term, $\textrm{Re}\left [ e^{i \phi N} \Phi(\phi) \right ]$.}.
While conceptually simple, measuring the real part of $\Phi(\phi)$ in this way requires the ability to reset an individual qubit during the execution of the protocol.
Alternatively, one can delay the projection to the end of the protocol by swapping in an ancilla qubit in the state $\ket{0}$ (taking ${V} = {\sigma}^z$), and post-selecting on the final state of the ancilla qubit.
%which may not be feasible for certain experimental platforms.

Our second approach, analogous to a standard Loschmidt echo, is to perform the full inverse of the state preparation procedure, as shown in Fig.~\ref{fig:opposite_quad}.  
This approach can be applied with either local or global controls, but, for specificity, let us focus on the variant with local controls.
The conceptually simplest quantity to analyze is the return probability, $P_0$, given by
\begin{equation} \label{eq:V_real}
\begin{split}
P_0 &= \frac 1 4 \left|\bra{\textbf{0}}(\mathbbm{1}-i V(t)) e^{-i\phi S_z} (\mathbbm{1}+i V(t))\ket{\textbf{0}}\right|^2 \\
    &= \frac 1 4 \left|e^{-i\phi \frac N 2}+\bra{\textbf{0}} V(t) e^{i\phi S_z} V(t) \ket{\textbf{0}}\right|^2 \\
    &= \frac 1 4 \left (1 + \left | \Phi(\phi) \right |^2 + 2 \textrm{Re}\left[ e^{-i \phi \frac N 2} \Phi(\phi)  \right] \right)
\end{split} .
\end{equation}
The real part of $\Phi(\phi)$ can easily be inferred by combining this quantity with the outcome from Eq.~(\ref{eq:im_Phi}).
% )
This approach directly generalizes to the case of global controls by replacing the local rotation with a global one, e.g.~$e^{i\epsilon S_x}$.
In either case, an additional many-body unitary (i.e.~2 copies of $U$ and $U^\dagger$) is required compared to the previous protocols. 

We note that, although measuring the return probability is straightforward to analyze, it is highly sensitive to readout errors.
In practice, therefore, it is  better to measure either the average polarization or polarization distribution of the final state, both of which display qualitatively similar behavior to the return probability.
In particular, if the external signal applies a relative phase between the two components of the butterfly state, $\ket{\textbf{0}}$ and $ V(t) \ket{\textbf{0}}$, then the polarization distribution of the final state features two peaks---a fully polarized state and a random state centered about zero polarization---with a relative height that oscillates as function of $\phi$.
Readout noise broadens the two peaks, but they remain extensively separated.

\bibliography{refs}